\documentclass[%
 reprint,
nofootinbib,
 amsmath,amssymb,
 aps,
]{revtex4-2}

\usepackage{graphicx}
\usepackage{dcolumn}
\usepackage{bm}
\usepackage{amsmath}	
\usepackage{amssymb}	
\usepackage{physics}
\usepackage{tabularx}
\usepackage{makecell}
\usepackage{booktabs}
\usepackage{natbib}
\usepackage{aas_macros}
\usepackage[dvipsnames]{xcolor}
\usepackage{gensymb}
\usepackage{lineno}
\usepackage[normalem]{ulem}
\usepackage{hyperref}
\usepackage{orcidlink}

\begin{document}

\preprint{APS/123-QED}

\title{Measurements of the Thermal Sunyaev-Zel'dovich Effect with ACT and DESI Luminous Red Galaxies}

\author{R. Henry Liu$^{1,2,3}$ \orcidlink{0000-0001-6079-5683}}
\email{rh_liu@berkeley.edu}
\author{Simone Ferraro$^{2,1,3}$}%
\author{Emmanuel Schaan$^{4}$}
\author{Rongpu Zhou$^{2}$}
\author{Jessica Nicole Aguilar$^{2}$}
\author{Steven Ahlen$^{5}$}
\author{Nicholas Battaglia$^{6,7}$}
\author{Davide Bianchi$^{8}$}
\author{David Brooks$^{9}$}
\author{Todd Claybaugh$^2$}
\author{Shaun Cole$^{10}$}
\author{William R. Coulton$^{11,12}$}
\author{Axel de la Macorra$^{13}$}
\author{Arjun Dey$^{14}$}
\author{Kevin Fanning$^{15,3}$}
\author{Jaime E. Forero-Romero$^{16,17}$}
\author{Enrique Gaztañaga$^{18,19,20}$}
\author{Yulin Gong$^{6}$}
\author{Satya Gontcho A Gontcho$^2$}
\author{Daniel Gruen$^{21,22}$}
\author{Gaston Gutierrez$^{23}$}
\author{Boryana Hadzhiyska$^{24, 2, 3}$}
\author{Klaus Honscheid$^{25,26,27}$}
\author{Cullan Howlett$^{28}$}
\author{Robert Kehoe$^{29}$}
\author{Theodore Kisner$^2$}
\author{Anthony Kremin$^2$}
\author{Aleksandra Kusiak$^{11,30}$}
\author{Andrew Lambert$^2$}
\author{Martin Landriau$^{2}$}
\author{Laurent Le Guillou$^{31}$}
\author{Michael Levi$^2$}
\author{Martine Lokken$^{32}$}
\author{Marc Manera$^{33, 32}$}
\author{Paul Martini$^{25, 34, 27}$}
\author{Aaron Meisner$^{14}$}
\author{Ramon Miquel$^{35, 32}$}
\author{Kavilan Moodley$^{36, 37}$}
\author{Jeffrey A. Newman$^{38}$}
\author{Gustavo Niz$^{39, 40}$}
\author{Nathalie Palanque-Delabrouille$^{41, 2}$}
\author{Will Percival$^{42,43,44}$}
\author{Francisco Prada$^{45}$}
\author{Ignasi Pérez-Ràfols$^{46}$}
\author{Bernardita Ried Guachalla$^{47, 15, 4}$}
\author{Graziano Rossi$^{48}$}
\author{Eusebio Sanchez$^{49}$}
\author{David Schlegel$^2$}
\author{Michael Schubnell$^{50, 51}$}
\author{Hee-Jong Seo$^{52}$}
\author{Crist\'obal Sif\'on$^{53}$}
\author{David Sprayberry$^{14}$}
\author{Gregory Tarl\'{e}$^{51}$}
\author{Eve M. Vavagiakis$^{54, 55}$}
\author{Benjamin Alan Weaver$^{14}$}
\author{Edward J. Wollack$^{56}$}
\author{Hu Zou$^{57}$}

\affiliation{
$^{1}$Department of Physics, University of California, Berkeley, CA, 94720, USA \\  
$^{2}$Lawrence Berkeley National Laboratory, 1 Cyclotron Road, Berkeley, CA 94720, USA \\
$^{3}$Berkeley Center for Cosmological Physics, Department of Physics, University of California, Berkeley, CA 94720, USA \\
$^{4}$SLAC National Accelerator Laboratory, 2575 Sand Hill Road, Menlo Park, California 94025, USA \\
$^{5}$Physics Dept., Boston University, 590 Commonwealth Avenue, Boston, MA 02215, USA \\
$^{6}$Department of Astronomy, Cornell University, Ithaca, NY 14853, USA \\
$^{7}$Université Paris Cité, CNRS, Astroparticule et Cosmologie, F-75013 Paris, France\\
$^{8}$Dipartimento di Fisica ``Aldo Pontremoli'', Universit\`a degli Studi di Milano, Via Celoria 16, I-20133 Milano, Italy  \\
$^{9}$Department of Physics \& Astronomy, University College London, Gower Street, London, WC1E 6BT, UK \\
$^{10}$Institute for Computational Cosmology, Department of Physics, Durham University, South Road, Durham DH1 3LE, UK \\
$^{11}$Kavli Institute for Cosmology Cambridge, Madingley Road, Cambridge CB3 0HA, UK \\
$^{12}$DAMTP, Centre for Mathematical Sciences, University of Cambridge, Wilberforce Road, Cambridge CB3 OWA, UK \\
$^{13}$Instituto de F\'{\i}sica, Universidad Nacional Aut\'{o}noma de M\'{e}xico,  Circuito de la Investigaci\'{o}n Cient\'{\i}fica, Ciudad Universitaria, Cd. de M\'{e}xico  C.~P.~04510,  M\'{e}xico \\
$^{14}$NSF NOIRLab, 950 N. Cherry Avenue, Tucson, AZ 85719, USA \\
$^{15}$Kavli Institute for Particle Astrophysics and Cosmology, Stanford University, Menlo Park, CA 94305, USA \\
$^{16}$Departamento de F\'isica, Universidad de los Andes, Carrera 1 No. 18A-10, Edificio Ip, CP 111711, Bogot\'a, Colombia \\
$^{17}$Observatorio Astron\'omico, Universidad de los Andes, Carrera 1 No. 18A-10, Edificio H, CP 111711 Bogot\'a, Colombia \\
$^{18}$Institut d'Estudis Espacials de Catalunya (IEEC), c/ Esteve Terradas 1, Edifici RDIT, Campus PMT-UPC, 08860 Castelldefels, Spain \\
$^{19}$Institute of Cosmology and Gravitation, University of Portsmouth, Dennis Sciama Building, Portsmouth, PO1 3FX, UK \\
$^{20}$Institute of Space Sciences, ICE-CSIC, Campus UAB, Carrer de Can Magrans s/n, 08913 Bellaterra, Barcelona, Spain \\
$^{21}$Excellence Cluster ORIGINS, Boltzmannstrasse 2, D-85748 Garching, Germany \\
$^{22}$University Observatory, Faculty of Physics, Ludwig-Maximilians-Universit\"{a}t, Scheinerstr. 1, 81677 M\"{u}nchen, Germany \\
$^{23}$Fermi National Accelerator Laboratory, PO Box 500, Batavia, IL 60510, USA \\
$^{24}$Miller Institute for Basic Research in Science, University of California, Berkeley, CA, 94720, USA \\
$^{25}$Center for Cosmology and AstroParticle Physics, The Ohio State University, 191 West Woodruff Avenue, Columbus, OH 43210, USA \\
$^{26}$Department of Physics, The Ohio State University, 191 West Woodruff Avenue, Columbus, OH 43210, USA \\
$^{27}$The Ohio State University, Columbus, 43210 OH, USA \\
$^{28}$School of Mathematics and Physics, University of Queensland, Brisbane, QLD 4072, Australia \\
$^{29}$Department of Physics, Southern Methodist University, 3215 Daniel Avenue, Dallas, TX 75275, USA \\
$^{30}$Institute of Astronomy, University of Cambridge, Madingley Road, Cambridge, CB3 0HA, UK \\
$^{31}$Sorbonne Universit\'{e}, CNRS/IN2P3, Laboratoire de Physique Nucl\'{e}aire et de Hautes Energies (LPNHE), FR-75005 Paris, France \\
$^{32}$Institut de F\'{i}sica d'Altes Energies (IFAE), The Barcelona Institute of Science and Technology, Campus UAB, 08193 Bellaterra (Barcelona) Spain\\
$^{33}$Departament de F\'{i}sica, Serra H\'{u}nter, Universitat Aut\`{o}noma de Barcelona, 08193 Bellaterra (Barcelona), Spain \\
$^{34}$Department of Astronomy, The Ohio State University, 4055 McPherson Laboratory, 140 W 18th Avenue, Columbus, OH 43210, USA \\
$^{35}$Instituci\'{o} Catalana de Recerca i Estudis Avan\c{c}ats, Passeig de Llu\'{\i}s Companys, 23, 08010 Barcelona, Spain \\
$^{36}$Astrophysics Research Centre, University of KwaZulu-Natal, Westville Campus, Durban 4041, South Africa \\
$^{37}$School of Mathematics, Statistics \& Computer Science, University of KwaZulu-Natal, Westville Campus, Durban 4041, South Africa \\
$^{38}$Department of Physics \& Astronomy and Pittsburgh Particle Physics, Astrophysics, and Cosmology Center (PITT PACC), University of Pittsburgh, 3941 O'Hara Street, Pittsburgh, PA 15260, USA \\
$^{39}$ Departamento de F\'{\i}sica, DCI-Campus Le\'{o}n, Universidad de Guanajuato, Loma del Bosque 103, Le\'{o}n, Guanajuato C.~P.~37150, M\'{e}xico \\
$^{40}$ Instituto Avanzado de Cosmolog\'{\i}a A.~C., San Marcos 11 - Atenas 202. Magdalena Contreras. Ciudad de M\'{e}xico C.~P.~10720, M\'{e}xico \\
$^{41}$IRFU, CEA, Universit\'{e} Paris-Saclay, F-91191 Gif-sur-Yvette, France\\
$^{42}$Department of Physics and Astronomy, University of Waterloo, 200 University Avenue W, Waterloo, ON N2L 3G1, Canada \\
$^{43}$Perimeter Institute for Theoretical Physics, 31 Caroline Street North, Waterloo, ON N2L 2Y5, Canada \\
$^{44}$Waterloo Centre for Astrophysics, University of Waterloo, 200 University Avenue W, Waterloo, ON N2L 3G1, Canada \\
$^{45}$Instituto de Astrof\'{i}sica de Andaluc\'{i}a (CSIC), Glorieta de la Astronom\'{i}a, s/n, E-18008 Granada, Spain \\
$^{46}$Departament de F\'isica, EEBE, Universitat Polit\`ecnica de Catalunya, c/Eduard Maristany 10, 08930 Barcelona, Spain \\
$^{47}$Department of Physics, Stanford University, Stanford, CA, USA 94305-4085 \\
$^{48}$Department of Physics and Astronomy, Sejong University, 209 Neungdong-ro, Gwangjin-gu, Seoul 05006, Republic of Korea \\
$^{49}$CIEMAT, Avenida Complutense 40, E-28040 Madrid, Spain \\
$^{50}$Department of Physics, University of Michigan, 450 Church Street, Ann Arbor, MI 48109, USA\\
$^{51}$University of Michigan, 500 S. State Street, Ann Arbor, MI 48109, USA \\
$^{52}$Department of Physics \& Astronomy, Ohio University, 139 University Terrace, Athens, OH 45701, USA \\
$^{53}$ Instituto de F\'isica, Pontificia Universidad Cat\'olica de Valpara\'iso, Casilla 4059, Valpara\'iso, Chile \\
$^{54}$ Department of Physics, Duke University, Durham, NC 27710,
USA \\
$^{55}$ Department of Physics, Cornell University, Ithaca,
NY 14853, USA \\
$^{56}$ NASA Goddard Spaceflight Center, 8800 Greenbelt Rd, Greenbelt, MD 20771, USA \\
$^{57}$National Astronomical Observatories, Chinese Academy of Sciences, A20 Datun Road, Chaoyang District, Beijing, 100012, P.R. China
}




\date{\today}

\begin{abstract}

Cosmic Microwave Background (CMB) photons scatter off the free-electron gas in galaxies and clusters, allowing us to use the CMB as a backlight to probe the gas in and around low-redshift galaxies. The thermal Sunyaev-Zel'dovich effect, sourced by hot electrons in high-density environments, measures the thermal pressure of the target objects, shedding light on halo thermodynamics and galaxy formation and providing a path toward understanding the baryon distribution around cosmic structures. We use a combination of high-resolution CMB maps from the Atacama Cosmology Telescope (ACT) and photometric luminous red galaxy (LRG) catalogues from the Dark Energy Spectroscopic Instrument (DESI) to measure the thermal Sunyaev-Zel'dovich signal in four redshift bins from $z=0.4$ to $z=1.2$, with a combined detection significance of 19$\sigma$ when stacking on the fiducial CMB Compton-$y$ map. We discuss possible sources of contamination, finding that residual dust emission associated with the target galaxies is important and limits current analyses. We discuss several mitigation strategies and quantify the residual modelling uncertainty. This work complements closely related measurements of the kinematic Sunyaev-Zel'dovich and weak lensing of the same galaxies.


\end{abstract}

\maketitle


\section{Introduction}
\label{sec:Intro}

The baryonic content of the universe is a key ingredient of current cosmological models. Measurements of the power spectrum of the Cosmic Microwave Background (CMB) indicate that baryons comprise roughly 5\% of the total energy content of the universe \cite{Planck:2018vyg}, or about 16\% of the total mass. However, observations of galaxies in the late-time universe comprise only approximately 10\% of the cosmological abundance of baryons \cite{Fukugita2004}.

Most of the remaining baryons are thought to be present in the warm-hot intergalactic medium (WHIM) as well as the circumgalactic medium (CGM) \cite{Cen2006, McQuinn:2015icp}. 
Locating these missing baryons and uncovering their properties will further improve our understanding of the processes involved in galaxy and large-scale structure formation. A better understanding of the missing baryons will also aid the calibration of baryon effects in weak lensing, one of the primary limitations in the cosmological interpretation of current and future lensing data.
One of the best tools for probing the baryon distribution is through measurements of the scattering of the Cosmic Microwave Background (CMB) photons with free electrons at lower redshift  \cite{Sunyaev1970, Sunyaev1972, Sunyaev1980a, Sunyaev1980b, Carlstrom2002}.

The CMB photons interact with free electrons in late-time galaxies and clusters through Thomson (or Compton) scattering, leaving imprints on the CMB known as the Sunyaev-Zel'dovich (SZ) effects \cite{Zeldovich1969, Ostriker1986}. These are usually classified based on the relative motion between the electrons and the CMB rest frame. 
The two main types of SZ effects are the thermal and kinematic SZ effects, respectively known as tSZ and kSZ.
The tSZ effect is the primary focus of this work, and arises from the scattering of CMB photons by the hot electrons in and around low-redshift halos. The size of the effect is proportional to both the electron number density and temperature, thus probing the \textit{thermal pressure} of the ionized gas.
The kSZ effect meanwhile is caused by the Doppler boosting of CMB photons due to the bulk motion of the gas, and is proportional to electron density (see \cite{Birkinshaw1999, Carlstrom2002, Mroczkowski2019, Bianchini:2025ksz} for reviews on the topic). As the magnitude of the SZ effect is independent of redshift, they are well suited for studies of high redshift galaxies and clusters.

Due to the fact that electron pressure is higher in more massive haloes, the tSZ signal scales as a higher power of halo mass ($\propto M^{5/3}$), and receives a large part of its contribution from the most massive objects in any sample. Measurements of the tSZ effect could be used to model gas temperature \cite{Pandey2019}, and clustering statistics when combined with lensing and gas emissions \cite{Shirasaki2020}.
Furthermore, measurements of the tSZ effect can be combined with measurements of the kSZ effect, and the
joint analysis of the two effects would allow for constraints on gas density and temperature, halo thermodynamics, galaxy feedback and formation, as well as baryon effects on lensing \cite{Battaglia2017}. For this reason, we perform the tSZ measurement on the same sample and in the same redshift bins as the kSZ measurements of \cite{Hadzhiyska2024, RiedGuachalla2025} to allow for these joint analyses. 

The use of the tSZ effect to probe the gas content of dark matter halos has a long history. In the last decade, maps from the Planck satellite \cite{Planck:2018nkj} have been used in combination with a variety of galaxy surveys to trace the hot gas in and around low redshift galaxies and clusters (see for example \cite{Kou:2022gcr, Ibitoye:2022oot, Greco:2014vwa, Yan:2021gfo, Chen:2022zii, Koukoufilippas:2019ilu}).
The arcminute angular resolution of ground-based wide-field CMB experiments such as the Atacama Cosmology Telescope (ACT, \cite{Fowler:2007dn}) and the South Pole Telescope (SPT, \cite{SPT:2018vbo}) permits the study of gas on smaller scales, closer to the galaxies' centres.

Data from ACT and Planck, together with galaxy catalogues from the Baryon Oscillation Spectroscopic Survey (BOSS, \cite{2013AJ....145...10D}) has been used to measure the stacked profiles of the thermal and kinematic SZ effects \cite{SchaanFerraro2021, Vavagiakis2021, Calafut2021, Mallaby-Kay2023}, and followed up with constraints on gas thermodynamics using these SZ measurements \cite{Amodeo2021}. Similar measurements have been obtained with galaxies from the Dark Energy Survey (DES, \cite{DES:2005dhi}), together with CMB data from ACT and Planck \cite{Pandey2019} or SPT and Planck \cite{DESSPT:2022xmu, Soergel2016_SPT}. 
Furthermore, ACT and Planck data have also been used to explore the filamentary structure of the cosmic web \cite{Hincks2022, Lokken2022, Isopi:2024umb, Hadzhiyska:2024ecq}. 

We note that in all cases, a correct interpretation of the signal requires modelling of the gas that is correlated with the galaxies in question but that belongs to other galaxies along the same line-of-sight, the so-called ``2-halo'' term \cite{Hill2018, Amodeo2021, Moser:2021llm, Hadzhiyska:2023cjj}.

The tSZ signal has also been used to identify galaxy clusters: the effect produces a signal that is straightforward to isolate and is independent of redshift\footnote{With some residual redshift dependence on the detection efficiency due to the redshift-dependent angular diameter in the sky, as well as a redshift dependence on the identification of optical counterparts.}.
SZ-selected cluster catalogues have been published from multiple collaborations \cite{Planck2013_XX, Planck:2015koh, Hilton2021, Bleem2024, Zubeldia2024}.

In this paper, we follow previous work closely \cite{SchaanFerraro2021} and combine data from the most recent Luminous Red Galaxy (LRG) photometric catalogue \cite{Zhou2023_LRGSample} from the Dark Energy Spectroscopic Instrument (DESI, \cite{DESICollaboration2016a}) and maps of the tSZ signal obtained from multi-frequency CMB maps from ACT DR6 and Planck \cite{Coulton2023_ACTDR6maps}. We stack the tSZ maps at the location of the LRG sample to measure the tSZ signal. We split the LRGs into four redshift bins spanning the range $z \approx 0.4 - 1.2$ and repeat the measurement as a function of radial distance from the centre of the galaxy (or ``aperture'' in what follows).

This paper is organized as follows: In Section \ref{sec:Theory} we will discuss the formalism behind the tSZ effect. Section \ref{sec:Data} will present the galaxy catalogues and the CMB maps used in our analysis. Section \ref{sec:Analysis} will discuss the methodology used in our tSZ stacking analysis Section \ref{sec:Results} will feature our results as well as a discussion of their meaning, before concluding in Section \ref{sec:Conclusion}.

\section{Theory}
\label{sec:Theory}

The tSZ effect is caused by the inverse Compton scattering of CMB photons interacting with free electrons of the hot ionized intergalactic gas. The random motions of the thermal electrons in the hot gas boost the energy of the scattered CMB photons, resulting in a characteristic distortion of the CMB frequency spectrum (see \cite{Carlstrom2002} for a review).

In particular, the tSZ effect leads to a spectral $y$-type distortion proportional to the square of the electron thermal velocity $v_{\mathrm{th}}$, which itself is proportional to the electron temperature $T_e$. 
The electron temperature is thus changed by \cite{Sunyaev1970, Sunyaev1972, Sunyaev1980a, Sunyaev1980b, Carlstrom2002}:
\begin{equation}
    \frac{\delta T_{\mathrm{tSZ}}(\hat{\mathbf{n}})}{T_\mathrm{CMB}} = f_{\mathrm{tSZ}}(\nu)y(\hat{\mathbf{n}}) \;.
    \label{eq:tSZ}
\end{equation}
Where $f_\mathrm{tSZ}$ is the frequency dependence given by
\begin{equation}
    f_\mathrm{tSZ} = x \coth(x/2) - 4 \;,
    \label{tSZ_SED}
\end{equation}
with $x=h\nu/k_BT_\mathrm{CMB}$, while the amplitude is given by the (dimensionless) Compton-$y$ parameter:
\begin{equation}
    y(\hat{\mathbf{n}}) = \frac{k_B\sigma_T}{m_e c^2} \int \frac{d\chi }{1+z} n_e(\chi \hat{\mathbf{n}}, z)T_e(\chi \hat{\mathbf{n}}) \;,
\end{equation}
with $\sigma_T$ as the Thomson cross section, $n_e$ as the free electron number density, $\chi$ as the comoving distance to redshift $z$, $m_e$ as the electron mass and $k_B$ as the Boltzmann constant. Note that the $y$-parameter is independent of frequency.

The particular frequency dependence of the signal outlined in Eq. \ref{tSZ_SED}, which is independent of electron temperature in the non-relativistic limit, allows for the creation of Compton-$y$ maps with unit response to the tSZ signal and reduced foregrounds by combining CMB maps at different frequencies \cite{Remazeilles:2010hq,Planck2015_XXII, Coulton2023_ACTDR6maps, McCarthy2024a}.

As measurements of the $y$-parameter are proportional to the product of the free electron density of the gas as well as the gas temperature, we can use these measurements to better understand halo thermodynamics \cite{Battaglia2017}, clustering filaments \cite{Lokken2022}, active galactic nuclei feedback and formation \cite{Grayson2023}, and baryon effects in lensing \cite{Amodeo2021, McCarthy:2024tvp, DES:2024iny}. Furthermore, combining measurements of the kSZ and tSZ effects of the same halos allows for model-independent measurements of the electron temperature as demonstrated on the BOSS sample in \cite{SchaanFerraro2021, Amodeo2021}.


\section{Data}
\label{sec:Data}

\subsection{DESI Survey Galaxy Sample}
\label{sec:DESI_Galaxies}

The Dark Energy Spectroscopic Instrument (DESI) is a robotic, fiber-fed, highly multiplexed spectroscopic surveyor that operates on the Mayall 4-meter telescope at Kitt Peak National Observatory \cite{DESICollaboration2022}, with the goal of determining the nature of dark energy through the most precise measurement of the expansion history of the universe ever obtained \cite{Snowmass2013}. DESI, which can obtain simultaneous spectra of almost 5000 objects over a $\sim 3\degree $ field \cite{DESICollaboration2016b, Silber2023, Miller2024}, is currently conducting a five-year survey of about a third of the sky. This campaign has obtained spectra for approximately 50 million galaxies and quasars \cite{DESICollaboration2016a}, following a survey validation campaign \cite{DESICollaboration2023a}, and an early data release (EDR) \cite{DESICollaboration2023b}.

Among the results using the First Data Release (DR1), are Key Papers presenting the two-point clustering measurements and validation \cite{DESI2024_Y1_II}, BAO measurements from galaxies and quasars \cite{DESI2024_Y1_III}, and from the Lyman-$\alpha$ forest \cite{DESI2024_Y1_IV}, as well as a full-shape study of galaxies and quasars \cite{DESI2024_Y1_V}. There are cosmological results from the BAO measurements \cite{DESI2024_Y1_VI} and the full-shape analysis \cite{DESI2024_Y1_VII}, as well as constraints on primordial non-Gaussianity \cite{Chaussidon:2024qni}.

We use the photometric Luminous Red Galaxies (LRG) sample from DESI \cite{Zou2017, Dey2019, Zhou2020, Zhou2023_TargetSelection, Zhou2023_LRGSample}, produced using an extensive survey reduction pipeline \cite{Guy2023, Schlafly2023}.
In particular, we use the sample described in  \cite{Zhou2023_LRGSample}, which provides photometric redshifts and spectroscopic calibration of the redshift distribution of the sample. The main LRG sample is split into four redshift bins based on photometric redshift measurements, as described in \cite{Zhou2023_LRGSample}. 

\begin{figure}
    \centering
    \includegraphics[width=\columnwidth]{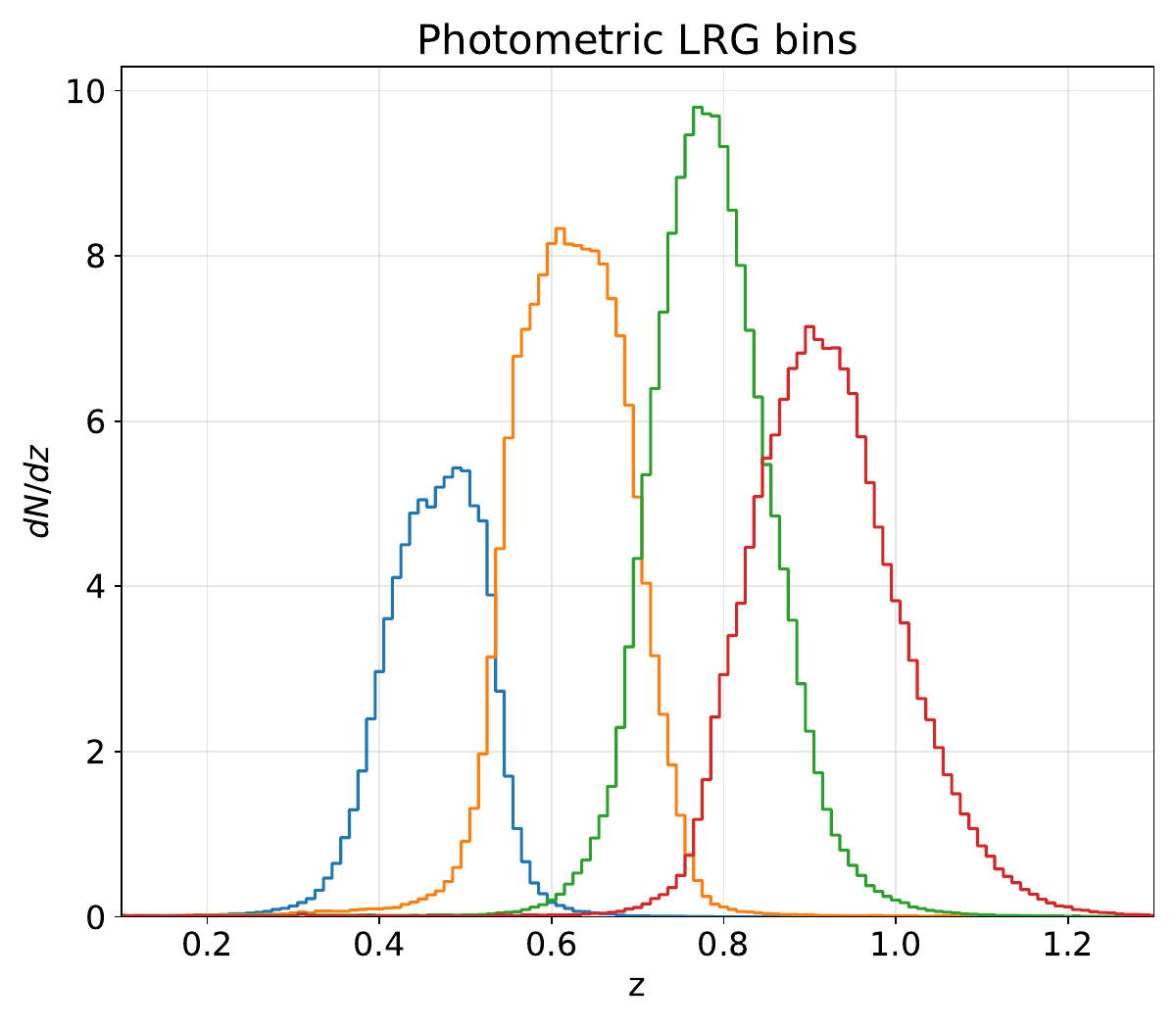}
    \caption{Spectroscopic distributions of four sub-sample photometric redshift bins, derived from 2.3 million DESI spectroscopic redshifts. The unit on the $y$ axis is the number of galaxies per square degree within the redshift bin, with width $dz=0.01$. 
    }
    \label{fig:DESI_redshiftbins}
\end{figure}

The redshift distributions of the four bins are shown in Figure \ref{fig:DESI_redshiftbins}, and report the number of galaxies in each bin in Table \ref{tab:DESI_LRG}.
We further note that the DESI main LRG sample catalogue is subject to a number of quality cuts from the full DESI catalogue, and is described in \cite{Zhou2023_LRGSample}. 

\begin{figure}
    \centering
    \includegraphics[width=\columnwidth]{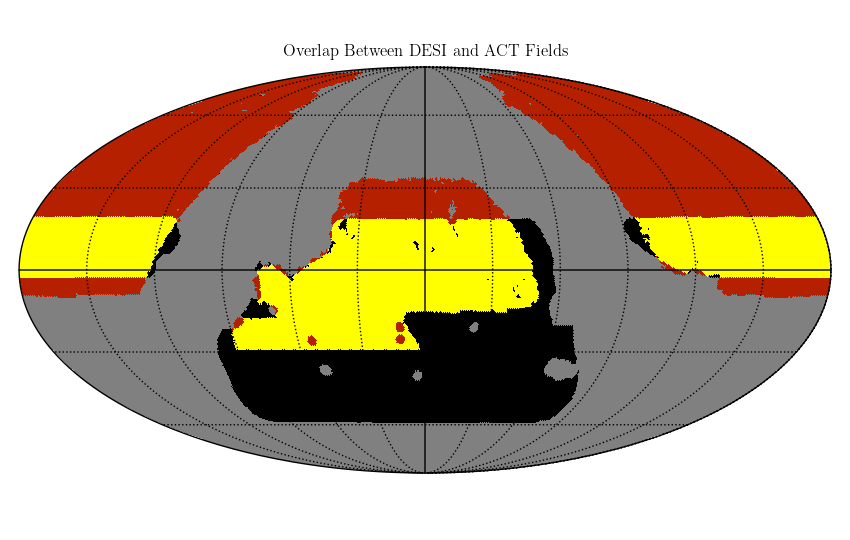}
    \caption{Overlap of ACT (black) and DESI (red) observation fields. The overlap area is shown in yellow, and represents the footprint overlap mask. The total overlap area is 7326 square degrees.}
    \label{fig:DESI_ACT_overlap}
\end{figure}


\begin{table*}
    \centering
    \begin{tabular}{|c|c|c|c|c|}
        \hline
        Bin & $\Bar{z}$ & $\Bar{n}_\theta$ & \# of objects in bin (overall) & \# of objects in bin (masked)\\
        \hline
        1 & 0.470 & 81.9 & 1,118,496 & 332,280\\
        2 & 0.628 & 148.1 & 2,031,303 & 608,100\\
        3 & 0.791 & 162.4 & 2,240,982 & 671,738\\
        4 & 0.924 & 148.3 & 2,049,158 & 615,543\\
        \hline
    \end{tabular}    
    \caption{Key information regarding the DESI LRG sample used; $\Bar{z}$ is the mean redshift of each bin, $\Bar{n}_\theta$ is the number density per square degree. We note that the reduction in objects after masking comes from reducing the overall DESI field to the smaller overlap field, as shown in Figure \ref{fig:DESI_ACT_overlap}.
    }
    \label{tab:DESI_LRG}
\end{table*}

Recent work based on the DESI One-Percent Survey has performed a comprehensive analysis of the DESI Halo Occupation Distribution \cite{Yuan2023_DESI_HOD}, giving us an approximate measurement of the host halo masses. They find a mean halo mass of $\log_{10}\Bar{M}_h = 13.40^{+0.02}_{-0.02}$ for LRGs in $0.4<z<0.6$, roughly corresponding to our first two redshift bins, and a lower mass of  $\log_{10}\Bar{M}_h = 13.24^{+0.02}_{-0.02}$ for LRGs at higher redshift $0.6<z<0.8$. 

\subsection{Cosmic Microwave Background Maps}

We utilise high-resolution and low-noise maps of the CMB from the Atacama Cosmology Telescope (ACT) for our analysis \cite{Coulton2023_ACTDR6maps}. 
Specifically, we make use of a set of component-separated Compton-$y$ maps, made using a combination of ACT and Planck data. The ACT DR6 data includes observations from 2017-2022, in three frequency bands: f090 (77–112 GHz), f150 (124-172 GHz), and f220 (182-277 GHz). The Planck data used originates from the NPIPE data described in \cite{Planck2020_LVII}. The single-frequency Planck data covers the whole sky, but only the area overlapping with ACT DR6 is used.

This is achieved through the use of a needlet decomposition and component separation pipeline, 
described in detail in \cite{Coulton2023_ACTDR6maps}. First, pre-processing is done on the data: Point sources are subtracted and the 70\% galactic mask from Planck is applied, and both the Planck and ACT maps are convolved to a 1.6 arcmin Full Width at Half Maximum (FWHM) Gaussian beam, which matches the ACT resolution. A Fourier space filter is further added in the pre-processing step to remove modes contaminated by scan-synchronous pickup. 
Next, the input maps are transformed to the wavelet frame. This is achieved by convolving the maps with generalized ``needlet'' kernels, implemented as a series of spherical harmonic transforms and filters. 
The transformed maps are then component-separated using the Needlet-frame Internal Linear Combination (NILC) method \cite{Delabrouille:2008qd}. This method, described in Section III. C of \cite{Coulton2023_ACTDR6maps}, combines all measurements at each needlet scale into a map of the Compton-$y$ parameter. An inverse needlet decomposition is then performed to transform the map back into a real space basis, and Fourier space filtering is done to replace missing modes.


As will be discussed in Section \ref{sec:Results}, foreground contamination is a major issue with the tSZ stacked profiles, especially due to the effects of dust contamination from the Cosmic Infrared Background (CIB). This contamination, previously reported in other studies \cite{Greco:2014vwa, SchaanFerraro2021} leads us to consider a number of ways to mitigate the effects of CIB emission using data beyond the fiducial component-separated Compton-$y$ parameter map from ACT.
We show that a simple deprojection of a fiducial CIB model leads to unstable results and strong dependence on the assumed model. Further moment deprojection largely mitigates this model dependence, as discussed in Section \ref{sec:CIB_deproject}.

As a consistency check, we also use single-frequency CMB temperature maps from ACT DR5 \cite{Naess2020_ACTDR5} with a modified filter. This is discussed in Appendix \ref{app:Single_Freq}. 



\section{Analysis}
\label{sec:Analysis}

\subsection{Filtering}
\label{subsec:Filtering}

Our pipeline is similar to what was used in previous analyses with ACT and BOSS data \cite{SchaanFerraro2021}. In particular, we define a set of Compensated Aperture Photometry (CAP) spatial filters with varying aperture radius $\theta_d$, centred around each galaxy in the catalogue. The output of the CAP filter on a $y$-parameter map is defined by
\begin{equation}
    y(\theta_d) = \int d^2\theta\ y W_{\theta_d}(\theta) \;,
\end{equation}
for different window functions $W_{\theta_d}(\theta)$.

We first consider a window function corresponding to a disk-ring filter $W_{\theta_d}(\theta)$, which is defined as
\begin{equation}
\label{eq:diskring}
    W_{\theta_d}(\theta) = 
    \begin{cases}
     1  & \theta < \theta_d \\
     -1 & \theta_d < \theta < \sqrt{2} \theta_d \\
     0  & \text{otherwise}
    \end{cases} \;.
\end{equation}

The CAP filter measures the integrated temperature fluctuation within a disc of radius $\theta_d$ and subtracts the same signal measured in a concentric ring of equal area surrounding the disc. Since $W_{\theta_d}$ averages to zero over the full range of $\theta$, fluctuations on scales larger than the filter are nulled. This compensation property allows longer-wavelength fluctuations than the filter size to be cancelled out after filtering, effectively reducing noise from degree-scale CMB fluctuations and reducing the correlation between different CAP filter sizes. The output of the filter decreases for smaller disc radii and converges to the cumulative gas pressure for larger radii, allowing each CAP filter radius to act as a band-pass filter on the temperature map before stacking. Consequently, CAP filter profiles can be interpreted as approximations of the cumulative gas pressure profile.



If the true tSZ profile was known a priori and we were only interested in measuring the amplitude of the known profile, a matched filter would be the minimum variance unbiased linear estimator. However, the tSZ profile is not known and the goal of this study is to measure this profile. Therefore we use a set of filters and we vary the size of $\theta_d$ between 1 and 6 arcmin, which corresponds to approximately 0.5-4 virial radii, the physical scales relevant for feedback and baryonic effects. 


\subsection{Masking}
\label{sec:Masking}

In addition to the signal from LRGs, the ACT DR6 $y$-parameter maps contain bright point sources and clusters.
The masking comprises three components: large clusters, point sources, and the footprint base mask (the latter of which is shown in Figure \ref{fig:DESI_ACT_overlap}), used to mask out the galactic plane and non-observed areas.

As a preprocessing step in the Needlet ILC pipeline, bright point sources were removed. This step allows for a lower noise and less contaminated Compton-$y$ map. Two different methods were used for removing the point sources: subtracting a model of the source and inpainting the region around the sources. Inpainting removes the signal from a larger region of the sky and is only used on sources that cannot be accurately subtracted - extended sources or those with SNR $>70$. Subtraction is used for all sources detected at SNR $>5$ in the individual frequency point source catalogues. See \cite{Coulton2023_ACTDR6maps} and \cite{Qu2024_ACTLensing} for more details on the subtraction and inpainting procedures. Regions impacted by the subtraction and inpainting can bias this analysis: inpainted regions contain a constrained Gaussian noise realization, rather than the true sky, and the subtraction inadvertently removes sky contributions beyond just the point source, as the model is built from the data itself. We therefore mask any inpainted pixels, with a disc of radius 6 arcmin around very bright point sources and a 10 arcmin disc around bright extended objects, and use a disc mask of radius 3 arcmin around all subtracted sources.

Very massive galaxy clusters dominate the Compton-$y$ map and their very bright signal can skew the stacks, described in the subsequent section. In this work we choose to mask the most massive clusters as it will reduce the variance in our measurement at minimal cost to the signal. We use a cluster catalogue derived from the ACT DR6 data and mask all clusters detected with SNR $>6$. This catalogue was constructed in a very similar manner to the DR5 catalogue described in \cite{Hilton2021}.  



\subsection{Stacking}

In order to obtain a combined (mean) tSZ measurement from the measured temperatures on individual galaxies, we stack the measured temperatures $\mathcal{T}_i(\theta_d)$ (measured at each specific $\theta_d$) to return the average stacked tSZ profile $\mathcal{T}(\theta_d)$. 
In principle, if the noise per object and aperture $\sigma^2_{i, \theta_d}$, is available and independent object by object, the optimal inverse-noise weighted mean is given by:
\begin{equation}
    \mathcal{T}(\theta_d) = \frac{\sum_i \mathcal{T}_i(\theta_d)/\sigma_{i, \theta_d}^2}{\sum_i 1/\sigma_{i, \theta_d}^2} \;.
\end{equation}
However, given the complexity of the noise post-NILC component separation and CIB deprojection, we adopt a uniform weighting and expect this to be very close to optimal given the modest noise variations across the map:
\begin{equation}
\label{eq:stackaverage}
    \mathcal{T}(\theta_d) = \frac{\sum_i \mathcal{T}_i(\theta_d)}{N} \;,
\end{equation}
where $N$ is the number of galaxies in each stacked catalogue.

An object is considered ``masked" if the aperture of the object at the largest radius hits the mask area. The object is removed from the stack and not considered for any aperture radius. As a result, we stack the same sample of objects for all apertures.

\subsection{Covariance Matrix Computation}
\label{sec:covariance_analysis}

Since filters of different apertures overlap, the covariance matrix will have non-negligible off-diagonal components which need to be taken into account when interpreting the signal. The covariance of different CAP filters can in principle be computed analytically from the power spectrum for a map with uniform noise. However, the depth of the ACT CMB maps is non-uniform and anisotropic, which complicates this simple calculation. 

Instead, we follow previous work \cite{SchaanFerraro2021} and use bootstrap re-sampling of individual galaxies to compute our fiducial covariance matrices.
We draw with repetition from the galaxy catalogue to generate a resampled galaxy catalogue, with the same number of objects. This re-sampled catalogue is then used to measure the stacked tSZ profiles, and this re-sampling process is repeated 10,000 times, and used to infer the tSZ covariance matrices. 

This method produces an unbiased estimate of the tSZ covariance, in the limit of independent noise realizations of galaxies. As discussed in \cite{SchaanFerraro2021}, this method may fail in case of a very high number density, where the assumption of independent noise from galaxy to galaxy no longer applies because of the spatial overlap of several of them. However, following Appendix D of \cite{SchaanFerraro2021}, we expect the bootstrap covariance used here to be accurate to $\sim$10\%, sufficient for our analysis.

\subsection{Impact of the Cosmic Infrared Background}
\label{sec:CIB_deproject}

Thermal dust emissions from DESI LRGs in our sample, as well as other galaxies along the same line-of-sight (which we consider as part of the CIB), can bias the tSZ signal obtained by the NILC technique. Mitigating this effect is one of the main goals of this paper.

One possible path towards reducing this contamination is using CIB-deprojected $y$-parameter maps from ACT \cite{Coulton2023_ACTDR6maps}. The CIB spectral energy distribution (SED) is assumed to be a modified blackbody
\begin{equation}
\label{eq:f_CIB}
    f_{\mathrm{CIB}}(\nu) = \frac{A \left(\frac{\nu}{\nu_0}\right)^{3+\beta}}{\exp \frac{h\nu}{k_B T_{\rm CIB}}-1} \left(\frac{dB(\nu, T)}{dT}\biggr\rvert_{T = T_{\mathrm{CMB}}}\right)^{-1} \;,
\end{equation}
where the default parameter values $\beta = 1.7$ and $T_{\rm CIB} = 10.70$ K characterize the mean CIB properties, while $\nu_0$ is a pivot frequency and $A$ is a normalization constant. The parameter values for $\beta$ and $T_{\rm CIB}$ are obtained by a fit to the CIB monopole at 217, 353, and 545 GHz, based on \cite{Planck2014_XXX_CIB}.

However, as discussed in \cite{Coulton2023_ACTDR6maps}, there are two caveats that may make CIB deprojection less effective. The first is that the CIB Spectral Energy Distribution (SED) is not well understood, and the deprojection of an inaccurate CIB SED could lead to a residual CIB signal in the tSZ stacks. Second, the contamination in the case of tSZ stacking is expected to be dominated by the dust emission from the galaxies in the samples themselves. This contamination can have different properties compared to the CIB as a whole, which is the average over many galaxies over a much broader redshift range, and more heavily weighted towards galaxies at $z \gtrsim 1$. Therefore, the mean CIB SED, even if known perfectly, might not accurately describe the thermal dust emission from the sample of interest. As we shall see in Section \ref{sec:Results}, the results from this simple deprojection are very sensitive to the assumed parameters, making it insufficient for our purposes. 



Due to the parameter sensitivity just mentioned, we also consider moment-deprojected $y$ maps, following the method of \cite{Chluba2017} and implemented in \cite{Coulton2023_ACTDR6maps}. This method involves deprojecting a second spectral template that represents the first-order Taylor expansion centred at the assumed spectral index $\beta$. These maps additionally deproject the derivative spectrum, given by 
\begin{equation}
    f_{\mathrm{CIB}-d\beta} (\nu) = \frac{A \ln{(\nu/\nu_0)}\left(\frac{\nu}{\nu_0}\right)^{3+\beta}}{\exp \frac{h\nu}{k_B T_{\rm CIB}}-1} \left(\frac{dB(\nu, T)}{dT}\biggr\rvert_{T = T_{\mathrm{CMB}}}\right)^{-1} .
\end{equation}
We note that this is equivalent to $f_{\mathrm{CIB}}(\nu) \ln{(\nu/\nu_0)}$.
By deprojecting both a fiducial CIB SED and the derivative of the SED with respect to $\beta$, we remove the first-order sensitivity to the assumed fiducial value.
We note however that each deprojection comes with an additional noise cost in the ILC map. This extra noise is visible in the rightmost panel of Figure \ref{fig:yparam_stackedmaps}, though we note that the effect is mitigated when computing radially averaged stacked profiles.

\section{Results}
\label{sec:Results}

\subsection{CIB Deprojected maps}

\begin{figure}
    \centering
    \includegraphics[width=\columnwidth]{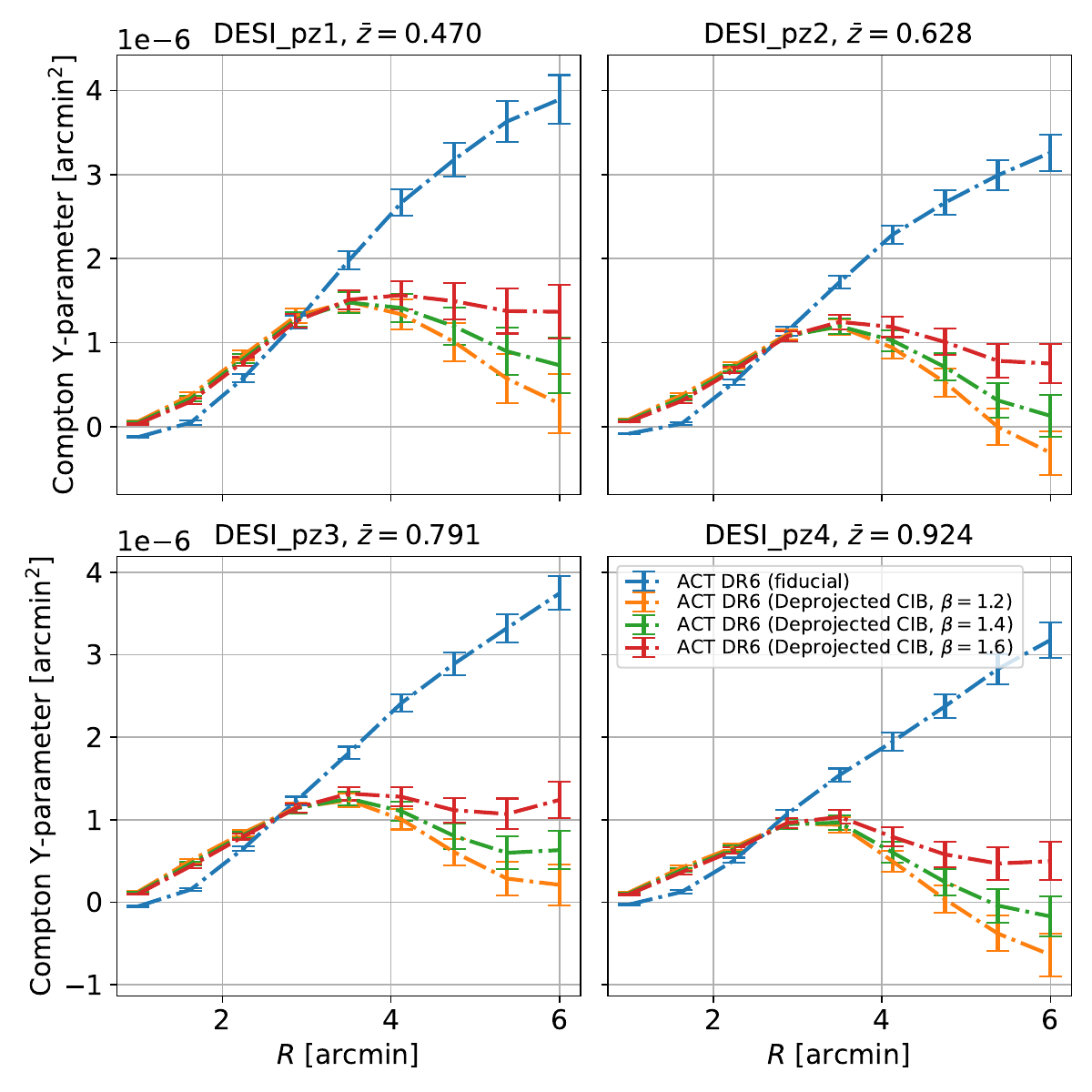}
    \caption{ACT DR6 fiducial $y$-parameter stacked profiles alongside stacked profiles using deprojected CIB $y$-parameter maps with varying values of the $\beta$ parameter. Each panel represents one photometric redshift bin as described in Section \ref{sec:DESI_Galaxies}.}
    \label{fig:yparam_stackedprofiles_cib}
\end{figure}

\begin{figure}
    \centering
    \includegraphics[width=\columnwidth]{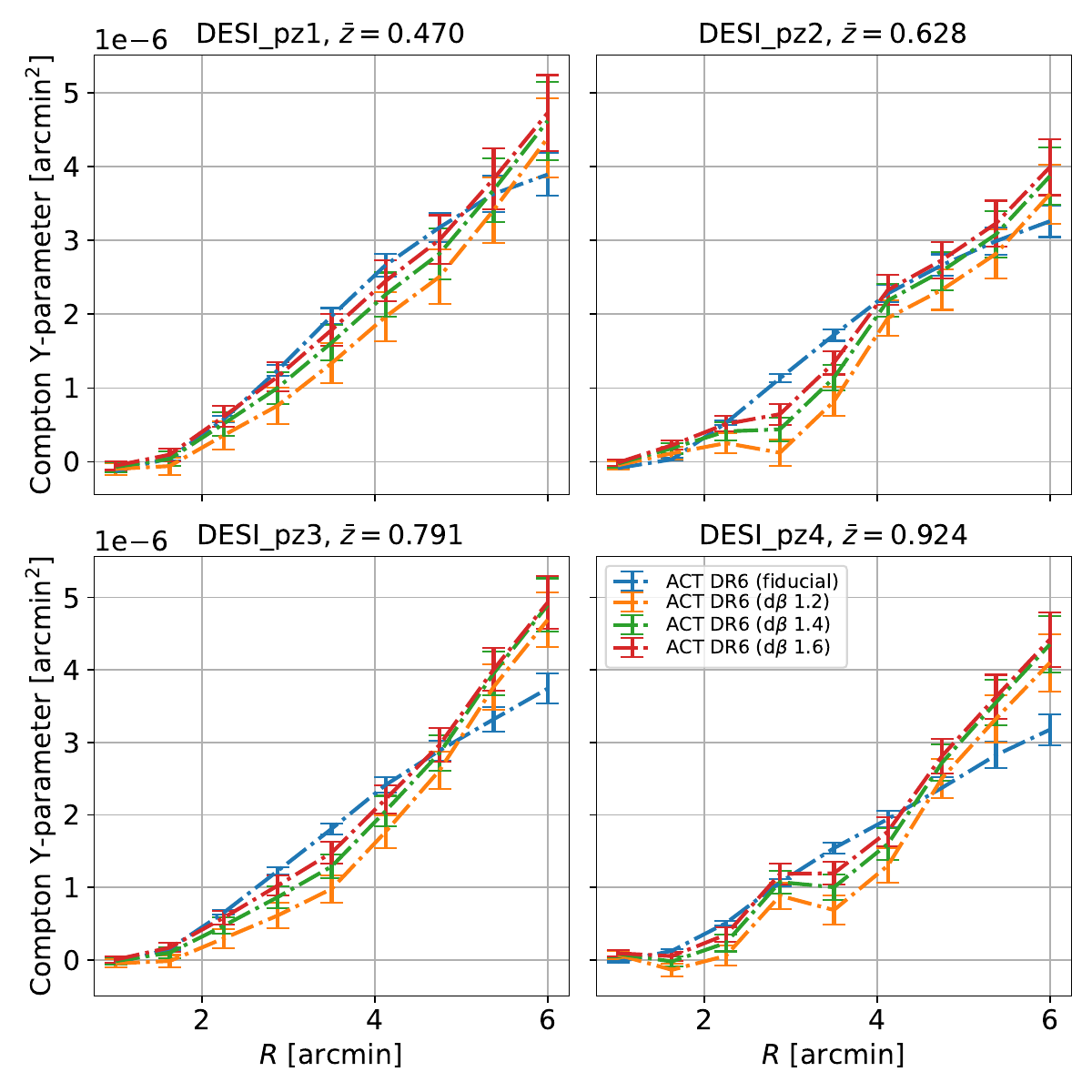}
    \caption{ACT DR6 fiducial $y$-parameter stacked profiles alongside the profiles stacked on $d\beta$ moment-deprojected, CIB-deprojected $y$-parameter maps, with varying values of $\beta$. }
    \label{fig:yparam_stackedprofiles_dbeta}
\end{figure}

We present the stacked tSZ profiles for the fiducial and CIB-deprojected $y$-parameter maps, as well as the $d\beta$ CIB moment-deprojected map, in Figures \ref{fig:yparam_stackedprofiles_cib} and \ref{fig:yparam_stackedprofiles_dbeta} respectively. These stacked profiles were generated using the disk-ring CAP filter as described in Section \ref{subsec:Filtering}. We show one stacked profile plot for each photo-$z$ bin in the DESI LRG catalogue. 
We see that the CIB-deprojected map deviates from the fiducial $y$-parameter map at large radii, as in Figure \ref{fig:yparam_stackedprofiles_cib}. Meanwhile, the stacked profiles for the fiducial and moment-deprojected CIB maps largely agree in value, especially in the lower photo-$z$ bins (see Figure \ref{fig:yparam_stackedprofiles_dbeta}). 

To compute the SNR, we fit the stacked profile with a 2-parameter Gaussian model and compute the $\chi^2_{\mathrm{model}}$. While not a physically-motivated model, it does provide a good fit to the data (as previously shown in \cite{SchaanFerraro2021}), and therefore can be used to quantify the SNR. In particular, we fit for both the width $\sigma$ and the amplitude $A$ of a Gaussian. The null $\chi^2_{\rm null}$ is computed by setting the profile to zero. 
The SNR and basic statistical quantities related to the fiducial stacking are reported in Table \ref{tab:SNR}. Fits to the tSZ and kSZ profiles informed by simulation as well as inference of physical properties will be the subject of an upcoming paper.

\begin{figure*}
    \centering
    \includegraphics[width=\textwidth]{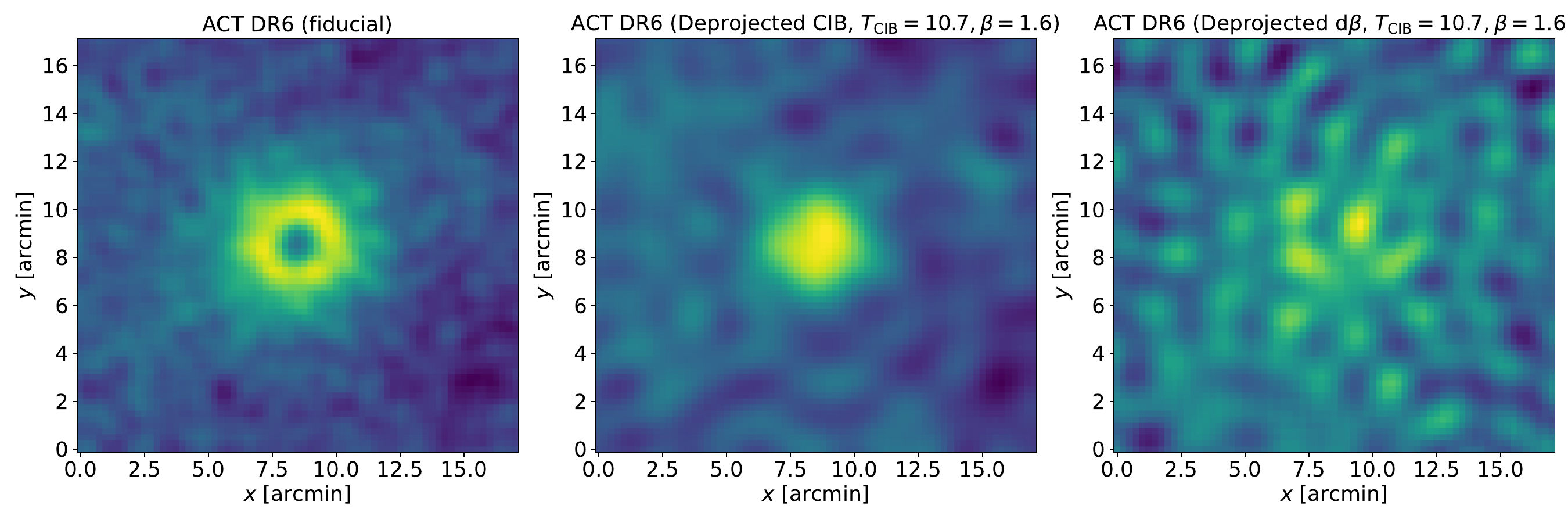}
    \caption{tSZ stacked maps for the fiducial $y$-parameter, CIB deprojected and $d\beta$ deprojected Compton-$y$ maps (at $T_{\rm CIB}=10.70$ K). These stacked maps are not generated using the CAP filter as in the case of the stacked profiles, but through the stacking of image cutouts themselves. Observe the higher noise and artifacts of the $\beta$ deprojected map. We note that through radial averaging, this result is mitigated and that the $d\beta$ deprojection is still capable of producing a signal. }
    \label{fig:yparam_stackedmaps}
\end{figure*}

We can easily observe the effect of dust contamination from the stacks fiducial Compton-$y$ parameter map, as shown in Figure \ref{fig:yparam_stackedmaps}: the fiducial Compton-$y$ parameter map has no deprojection of CIB, and we thus observe a central dust feature in the form of a negative signal in the stacked profiles in Figures \ref{fig:yparam_stackedprofiles_cib} and \ref{fig:yparam_stackedprofiles_dbeta} as well as in the centre of the panel on Figure \ref{fig:yparam_stackedmaps}. The significance of the CIB contamination in the fiducial map limits the interpretability of the signal, especially at small apertures.

The creation of CIB-deprojected $y$-parameter maps is described in detail in \cite{Coulton2023_ACTDR6maps}. As described in \cite{Coulton2023_ACTDR6maps}, the default  CIB parameters adopted are $\beta=1.7$ and $T_{\rm CIB}=10.7$ K, obtained from a preliminary version of the analysis conducted in \cite{McCarthy2024a, McCarthy2024b}. In our analysis, we fix $T_{\rm CIB}=10.7$ K while varying the values of $\beta$ as $1.2,\, 1.4,\,$ and $1.6$ in Figure \ref{fig:yparam_stackedprofiles_cib}. We see that changes in the CIB deprojection parameters result in large deviations in the final stacked tSZ profile.
The large shifts of the resulting tSZ profiles show the lack of stability of the CIB deprojection maps to changes in the CIB parameters and hint at large modelling uncertainty using this method.

Figure \ref{fig:yparam_stackedprofiles_dbeta} shows the tSZ signal with a deprojected CIB and the first moment in $\beta$, as described in Section \ref{sec:CIB_deproject}. We see that the deprojection of the first moment yields a more stable signal, and converges closer to the fiducial $y$-parameter map at large apertures. There is a residual scatter of order $\sim 20 \%$, which can be taken to represent the residual modelling uncertainty when deprojecting the CIB from ACT + Planck data alone. Of course, this uncertainty could be further reduced by the use of external datasets or future observations especially at higher frequencies. The corresponding plots for a different choice of fiducial CIB temperature ($T_{\rm CIB} = 24$ K) are shown in Appendix \ref{App:B}, and are in good agreement with the results presented here.




In addition to the radial stacked profiles shown in Figures \ref{fig:yparam_stackedprofiles_cib} and \ref{fig:yparam_stackedprofiles_dbeta}, we show the stacked tSZ maps in Figure \ref{fig:yparam_stackedmaps} for illustration. 
These stacked maps were created by stacking and averaging the maps following Eq. \ref{eq:stackaverage}, rather than using the CAP filters. As a result, the stacked maps are visually different from the stacked profiles as in Figure \ref{fig:yparam_stackedprofiles_cib}, where a CAP filter has been applied. We make note that the CIB dust contamination and central stacked map feature shown in these figures was also present in previous work involving ACT CMB maps \cite{SchaanFerraro2021, Amodeo2021, Vavagiakis2021}. 

The dust emission caused by the CIB is noticeable in the centre of the fiducial $y$-parameter stacked map, but not present in the other two. We do note that the $d\beta$ moment-deprojected map is more noisy: this is expected, because each deprojection comes at an increased noise cost, as discussed in \cite{Coulton2023_ACTDR6maps}. We note however that the radially-averaged and CAP-filtered profiles from these moment-deprojected maps are detected at high significance, as in Figure \ref{fig:yparam_stackedprofiles_dbeta}. The CIB deprojected map comes at a $\approx 10\%$ noise cost, while the $d\beta$-deprojected stacks have about $30 \%$ larger noise.




\subsection{Correlation Matrices}

\begin{figure}
    \centering
    \includegraphics[width=\columnwidth]{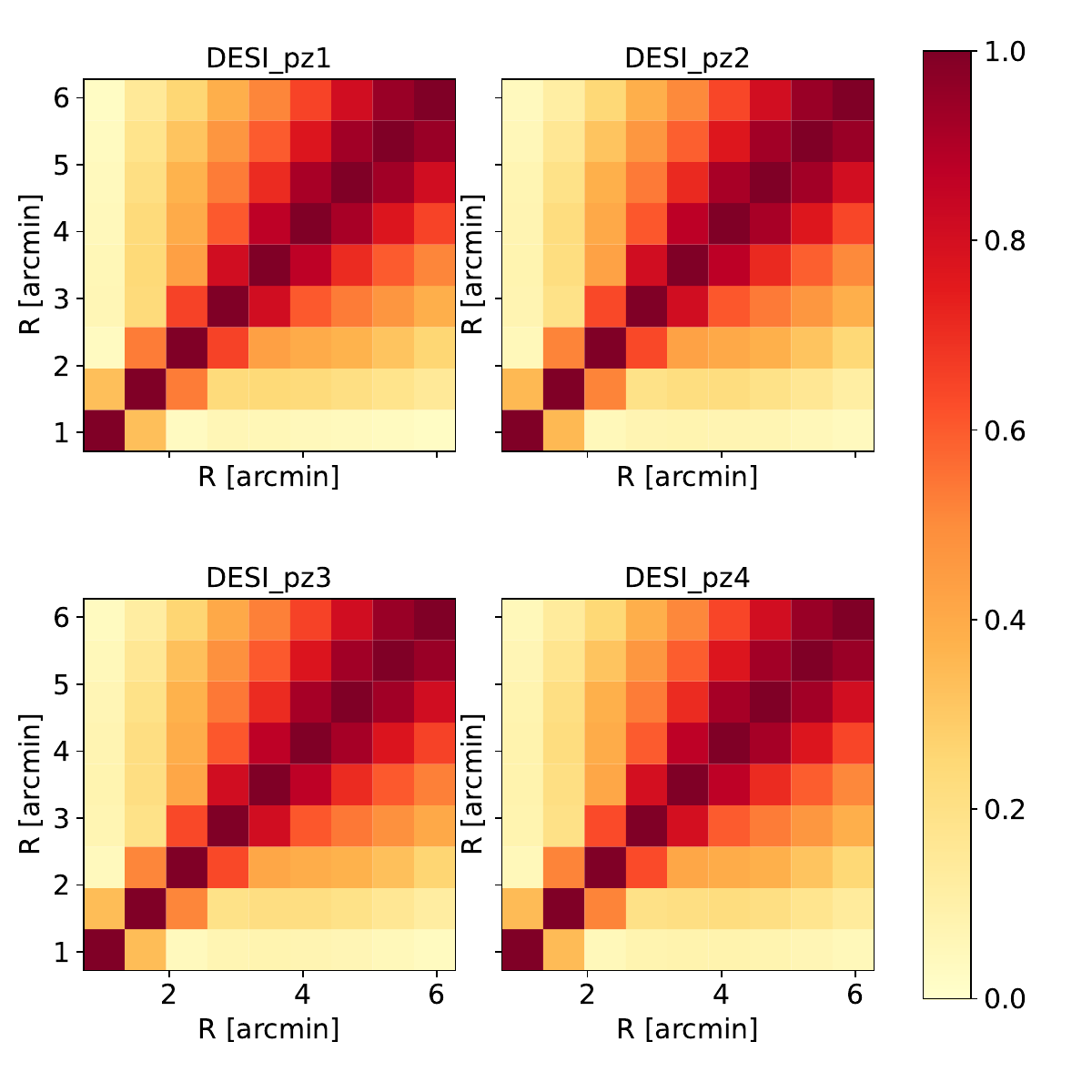}
    \caption{Correlation Matrix between the different CAP filters for the stacked profile from the ACT DR6 fiducial $y$-parameter map.}
    \label{fig:covariance_matrix}
\end{figure}

In addition to the radially stacked profile plots shown in the previous section, we also show the correlation matrices, obtained through bootstrap resampling of the galaxies as described in Section \ref{sec:covariance_analysis}. We show the resulting correlation matrix for the fiducial $y$-parameter map in Figure \ref{fig:covariance_matrix}. 


\begin{table}
    \centering
    \begin{tabular}{|c|c|c|c|c|c|}
    \hline
     Map & DESI Bin& $\chi^2_{\rm null}$ & $\chi^2_{\rm model}$ & dof &$\mathrm{SNR}_\mathrm{model}$ \\
    \hline
    \hline
     ACT DR6 (fiducial) & pz1 &  384.2 & 6.94 & 9 & 19.4\\
     ACT DR6 (fiducial) & pz2 &  567.2 & 16.4 & 9 & 23.5\\
     ACT DR6 (fiducial) & pz3 &  790.4 & 27.3 & 9 & 27.6\\
     ACT DR6 (fiducial) & pz4 &  508.1 & 18.6 & 9 & 22.1\\
    \hline
    \end{tabular}
    \caption{$\chi^2$ and signal to noise of the tSZ measurements on the fiducial $y$-parameter map with no deprojections. The $\mathrm{SNR}_\mathrm{model} = \sqrt{\chi^2_{\rm null} - \chi^2_{\rm model}}$ is computed with respect to the best-fit Gaussian profile, which provides a reasonable (even though not necessarily physical) fit to the data. 
    }
    \label{tab:SNR}
\end{table}

\section{Comparison to hydrodynamical simulations}

\begin{figure}
    \centering
    \includegraphics[width=\columnwidth]{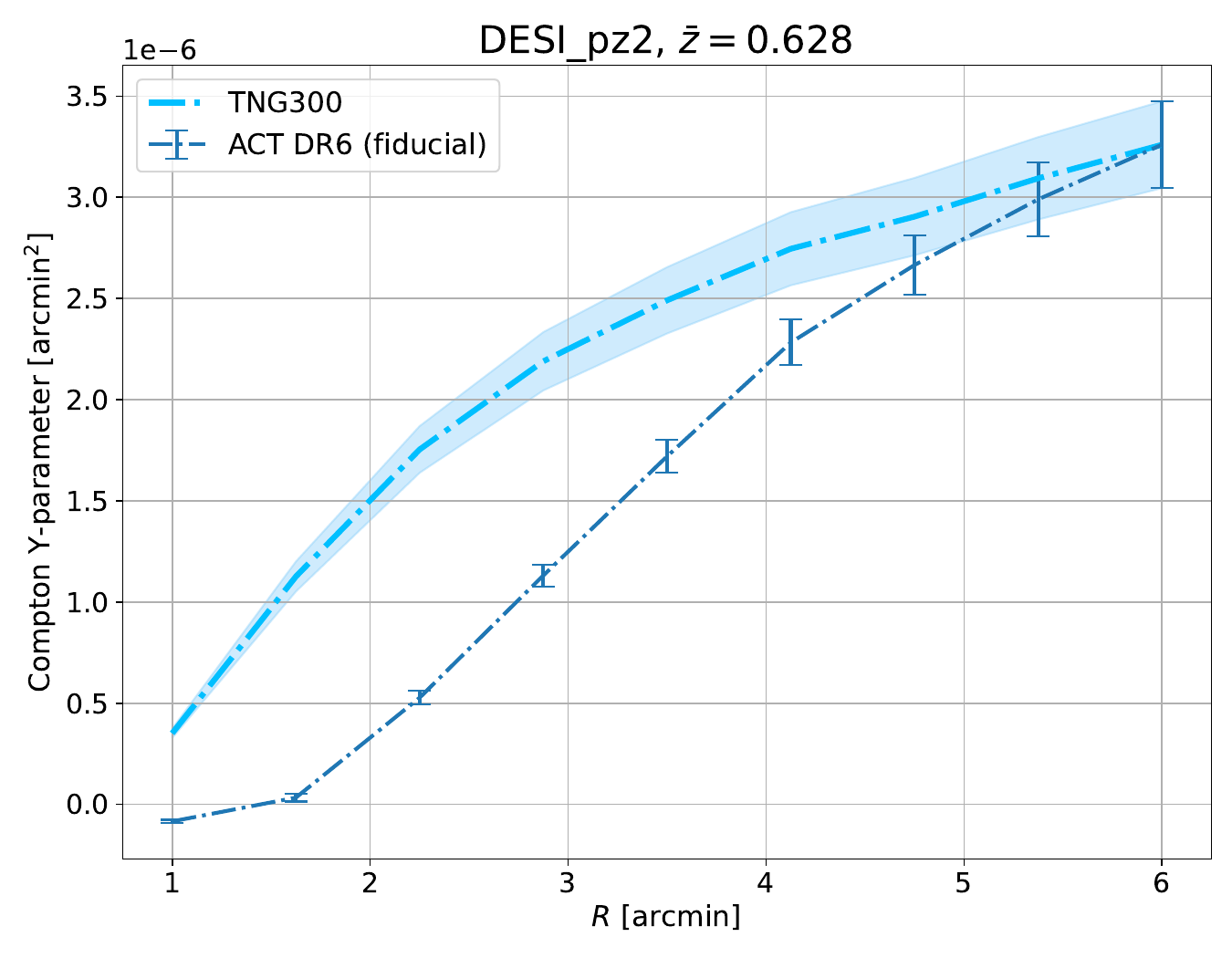}
    \caption{Comparison of the simulation-stacked profile with data in the DESI pz2 bin. }
    \label{fig:sim_comparison}
\end{figure}
The primary goal of this work is to present robust measurements of the tSZ signal from DESI LRG galaxies and to discuss possible sources of contamination. A full and fair comparison with hydrodynamical simulations and robust physical inference from it is a complex task that will be tackled in upcoming work.
Nonetheless, it is interesting to compare to a fiducial DESI LRG selection in the commonly-used IllustrisTNG simulation, similarly to what was performed in \cite{Hadzhiyska2024}. We project the simulations at a target redshift (in our case, $z=0.5$) onto a 2D plane and convolve the resulting profile with a $1.6'$ beam to mimic the resolution of the ACT maps. We then perform the stacking on this projected simulation map, and compare the results to our fiducial measurements from Figure \ref{fig:yparam_stackedprofiles_cib}. The resulting profile comparison is shown in Figure \ref{fig:sim_comparison}.

One notable difficulty in this comparison is the selection of halo masses. Following convention, we select haloes for LRGs for the range $10^{12.5} M_{\odot} \leq M_h \leq 3.8\times 10^{14} M_{\odot}$, with the lower bound being chosen for the conventional minimum LRG mass while the upper bound chosen from the estimated minimum mass of the masked clusters from \cite{Hilton2021}, as discussed in Section \ref{sec:Masking}. However, we note that the true mass distribution of our stacked sample is unknown. The steep dependence of the tSZ signal on mass further complicates the comparison. 

In the absence of truly reliable halo mass estimates for our LRG sample, we choose to match the amount of gas measured at large aperture between the simulations and the data (within the error bars of the data), ensuring that we are comparing galaxies with the same gas content (and therefore likely mass). This assumes that most of the gas associated with the LRGs is contained within our largest aperture (which corresponds to over 3 virial radii). This assumption holds true in most feedback models, and the case where this approach might fail is if feedback is so extreme that a large fraction of the gas has been completely expelled and is now unbound: we don't think this scenario is likely or supported by observations. Proper mass calibration with external data (for example, through lensing) will be able to answer this question definitively and won't require this matching at large aperture.


With the caveats noted above, we observe a large deficit of gas at $R < 3$ arcmin, suggestive that feedback in these galaxies is stronger than in the IllustrisTNG simulation, fully consistent with what was found for kSZ measurements of the same sample \cite{Hadzhiyska2024, RiedGuachalla2025}.

As a further confirmation of this statement, we can completely free up the amplitude (rather than matching it at large aperture). In this case, we still find that the best-fit
to all of the data points has $\chi^2_{bf} = 983$ for 8 degrees of freedom, which is an extremely poor fit, indicating a clear mismatch between the simulations and the data, independent of the amplitude. In particular, the simulation profile is \textit{too shallow} at intermediate scales, hinting at too little feedback. 


We conclude that the data hints at large amounts of feedback and a small gas fraction in the inner parts of the halo, in accordance with kSZ measurements. Proper calibration of the mean mass (and mass distribution) of the sample, marginalization over the satellite fraction (here fixed at the fiducial value), and a proper treatment of the selection function will be required for firmer conclusions, and will be subject to future work. This preliminary comparison, though, shows that the statistical quality of the data should allow for very interesting constraints on the feedback strength when the sample selection is properly addressed.





\section{Discussion and Conclusion}
\label{sec:Conclusion}

In this paper, we have measured the stacked profiles of the thermal Sunyaev-Zel'dovich effect around DESI Luminous Red Galaxies. Using Compensated Aperture Photometry filters, as well as the high-resolution ACT DR6 maps and the high-quality DESI LRG dataset, we have provided one of the highest significance measurements to date of the tSZ signal. 

As discussed in the text, one of the main challenges is dealing with the contamination from the Cosmic Infrared Background. In order to further analyze the effects of the CIB, we have also made measurements using a series of CIB-deprojected Compton-$y$ parameter maps, as well as $d\beta$ moment-deprojected maps. We were able to conclude that the CIB $d\beta$ moment-deprojected maps would allow us to mitigate the effects of dust while reducing the modelling uncertainty to $\sim 10-20\%$. A full CIB deprojection for our redshift range of interest may only be possible with the use of external data such as dust observations from PRIMA \cite{Moullet2023_PRIMA} or CCAT \cite{CCAT2023}, as well as data from Herschel (as used in \cite{Amodeo2021}) or observations from the James Webb Space Telescope. 


The measurement of the tSZ signal is an additional tracer of baryonic matter that could be used to constrain galaxy halo thermodynamics \cite{Battaglia2017}, filamentary structure \cite{Lokken2022}, galaxy feedback and formation \cite{Grayson2023}, baryon effects on lensing \cite{McCarthy2024b}, as well as the gas temperature when combined with similar kSZ measurements \cite{Amodeo2021}. 
Similar stacking work on the same samples to measure the kSZ signal with ACT and DESI data is in progress \cite{Hadzhiyska2024, RiedGuachalla2025}. 

In upcoming work \cite{Popik2025}, we will compare these measurements to the kSZ signal and to simulations while explicitly including the effect of satellite galaxies, miscentering and detailed modelling of the selection function and mass distribution of the sample. This will allow us to obtain the temperature profile of the gas within DESI galaxies, and provide model-specific constraints on the thermal energy injected into these galaxies through feedback as well as the amount of non-thermal pressure support present. Combined with weak lensing measurements, this will enable us to derive the complete thermodynamic properties of the halo, providing a comprehensive view of the relationship between visible and dark matter. Additionally, it will shed light on the complex processes involved in galaxy formation and evolution.


\section*{Data Availability}
Data points for the figures are available in digital format at \url{https://zenodo.org/records/14706729}.

\acknowledgments

R.H.L. is supported by the Postgraduate-Doctoral Scholarship from the Natural Sciences and Engineering Research Council of Canada (NSERC), funding reference number PGSD-567923-2022.
S.F. and R.Z. are supported by Lawrence Berkeley National Laboratory and the Director, Office of Science, Office of High Energy Physics of the U.S. Department of Energy under Contract No.\ DE-AC02-05CH11231. The work of E.S. received support from the U.S. Department of Energy under contract number DE-AC02-76SF00515 to SLAC National Accelerator Laboratory.
This research has made use of NASA's Astrophysics Data System and the arXiv preprint server.

This material is based upon work supported by the U.S. Department of Energy (DOE), Office of Science, Office of High-Energy Physics, under Contract No. DE-AC02-05CH11231, and by the National Energy Research Scientific Computing Center, a DOE Office of Science User Facility under the same contract. Additional support for DESI was provided by the U.S. National Science Foundation (NSF), Division of Astronomical Sciences under Contract No. AST-0950945 to the NSF's National Optical-Infrared Astronomy Research Laboratory; the Science and Technologies Facilities Council of the United Kingdom; the Gordon and Betty Moore Foundation; the Heising-Simons Foundation; the French Alternative Energies and Atomic Energy Commission (CEA); the National Council of Science and Technology of Mexico (CONACYT); the Ministry of Science and Innovation of Spain (MICINN), and by the DESI Member Institutions: \url{https://www.desi.lbl.gov/collaborating-institutions}.

The DESI Legacy Imaging Surveys consist of three individual and complementary projects: the Dark Energy Camera Legacy Survey (DECaLS), the Beijing-Arizona Sky Survey (BASS), and the Mayall $z$-band Legacy Survey (MzLS). DECaLS, BASS and MzLS together include data obtained, respectively, at the Blanco telescope, Cerro Tololo Inter-American Observatory, NSF's NOIRLab; the Bok telescope, Steward Observatory, University of Arizona; and the Mayall telescope, Kitt Peak National Observatory, NOIRLab. NOIRLab is operated by the Association of Universities for Research in Astronomy (AURA) under a cooperative agreement with the National Science Foundation. Pipeline processing and analyses of the data were supported by NOIRLab and the Lawrence Berkeley National Laboratory. Legacy Surveys also uses data products from the Near-Earth Object Wide-field Infrared Survey Explorer (NEOWISE), a project of the Jet Propulsion Laboratory/California Institute of Technology, funded by the National Aeronautics and Space Administration. Legacy Surveys was supported by: the Director, Office of Science, Office of High Energy Physics of the U.S. Department of Energy; the National Energy Research Scientific Computing Center, a DOE Office of Science User Facility; the U.S. National Science Foundation, Division of Astronomical Sciences; the National Astronomical Observatories of China, the Chinese Academy of Sciences and the Chinese National Natural Science Foundation. LBNL is managed by the Regents of the University of California under contract to the U.S. Department of Energy. The complete acknowledgments can be found at \url{https://www.legacysurvey.org/}.

Any opinions, findings, and conclusions or recommendations expressed in this material are those of the author(s) and do not necessarily reflect the views of the U. S. National Science Foundation, the U. S. Department of Energy, or any of the listed funding agencies.

The authors are honored to be permitted to conduct scientific research on Iolkam Du'ag (Kitt Peak), a mountain with particular significance to the Tohono O'odham Nation.

Support for ACT was through the U.S.~National Science Foundation through awards AST-0408698, AST-0965625, and AST-1440226 for the ACT project, as well as awards PHY-0355328, PHY-0855887 and PHY-1214379. Funding was also provided by Princeton University, the University of Pennsylvania, and a Canada Foundation for Innovation (CFI) award to UBC. ACT operated in the Parque Astron\'omico Atacama in northern Chile under the auspices of the Agencia Nacional de Investigaci\'on y Desarrollo (ANID). The development of multichroic detectors and lenses was supported by NASA grants NNX13AE56G and NNX14AB58G. Detector research at NIST was supported by the NIST Innovations in Measurement Science program. Computing for ACT was performed using the Princeton Research Computing resources at Princeton University, the National Energy Research Scientific Computing Center (NERSC), and the Niagara supercomputer at the SciNet HPC Consortium. SciNet is funded by the CFI under the auspices of Compute Canada, the Government of Ontario, the Ontario Research Fund–Research Excellence, and the University of Toronto. We thank the Republic of Chile for hosting ACT in the northern Atacama, and the local indigenous Licanantay communities whom we follow in observing and learning from the night sky.


\bibliographystyle{prsty.bst}

\bibliography{refs}

\newpage
\appendix

\section{Non-Cumulative tSZ Measurements}
\label{app:noncumulative}

So far, our tSZ measurements made use of cumulative Compensated Aperture Photometry (CAP) filters to measure the total signal within a particular radius. While useful, this cumulative measurement yields high correlations between the tSZ bins, as we can see in Figure \ref{fig:covariance_matrix}, making it difficult to study the tSZ signal at a particular radius independent of the inner or outer points.

\begin{figure}
    \centering
    \includegraphics[width=\linewidth]{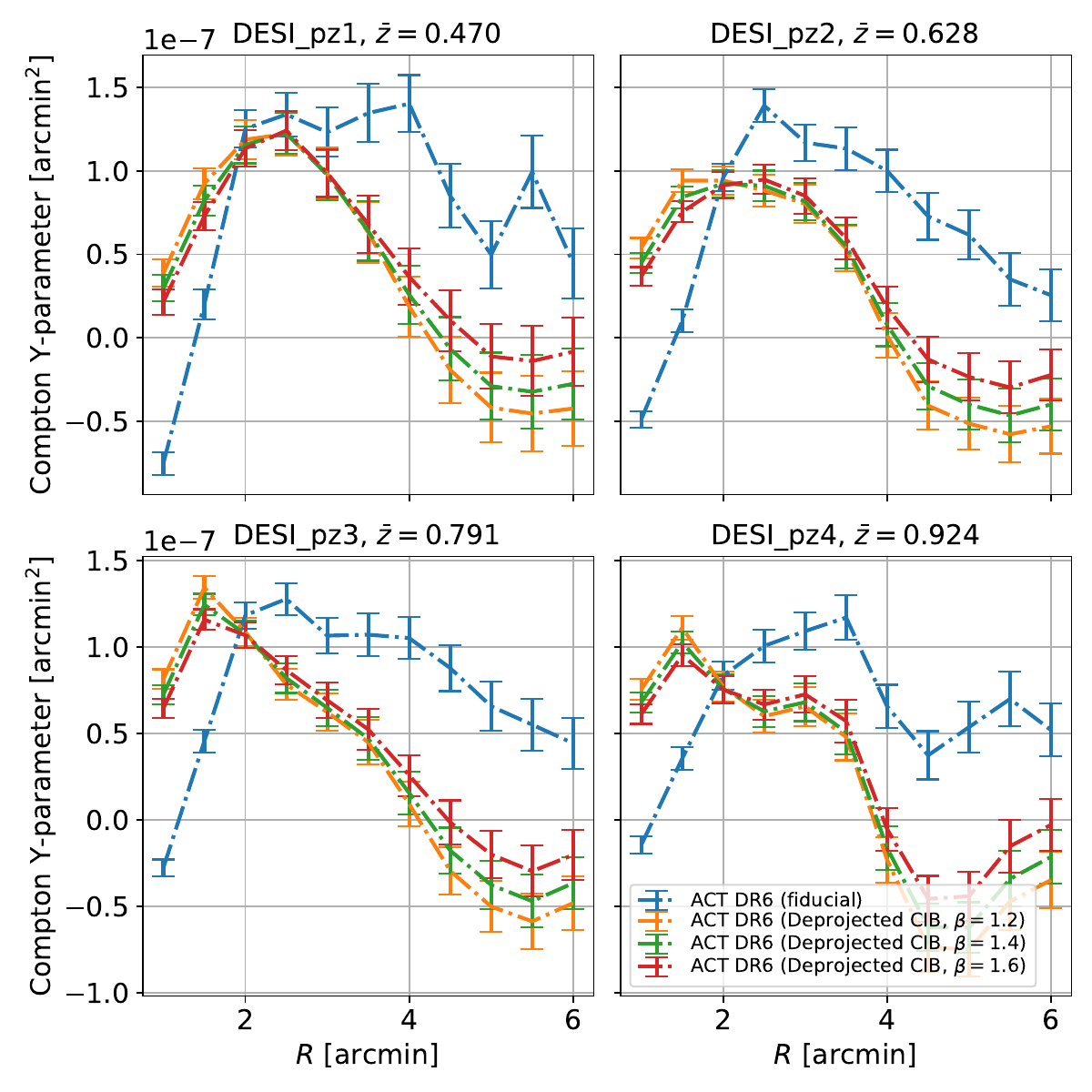}
    \caption{Non-Cumulative variant of Figure \ref{fig:yparam_stackedprofiles_cib} showing the stacked profiles for the fiducial and CIB-deprojected $y$-parameter maps.}
    \label{fig:stackedProfiles_noncum}
\end{figure}

\begin{figure}
    \centering
    \includegraphics[width=\linewidth]{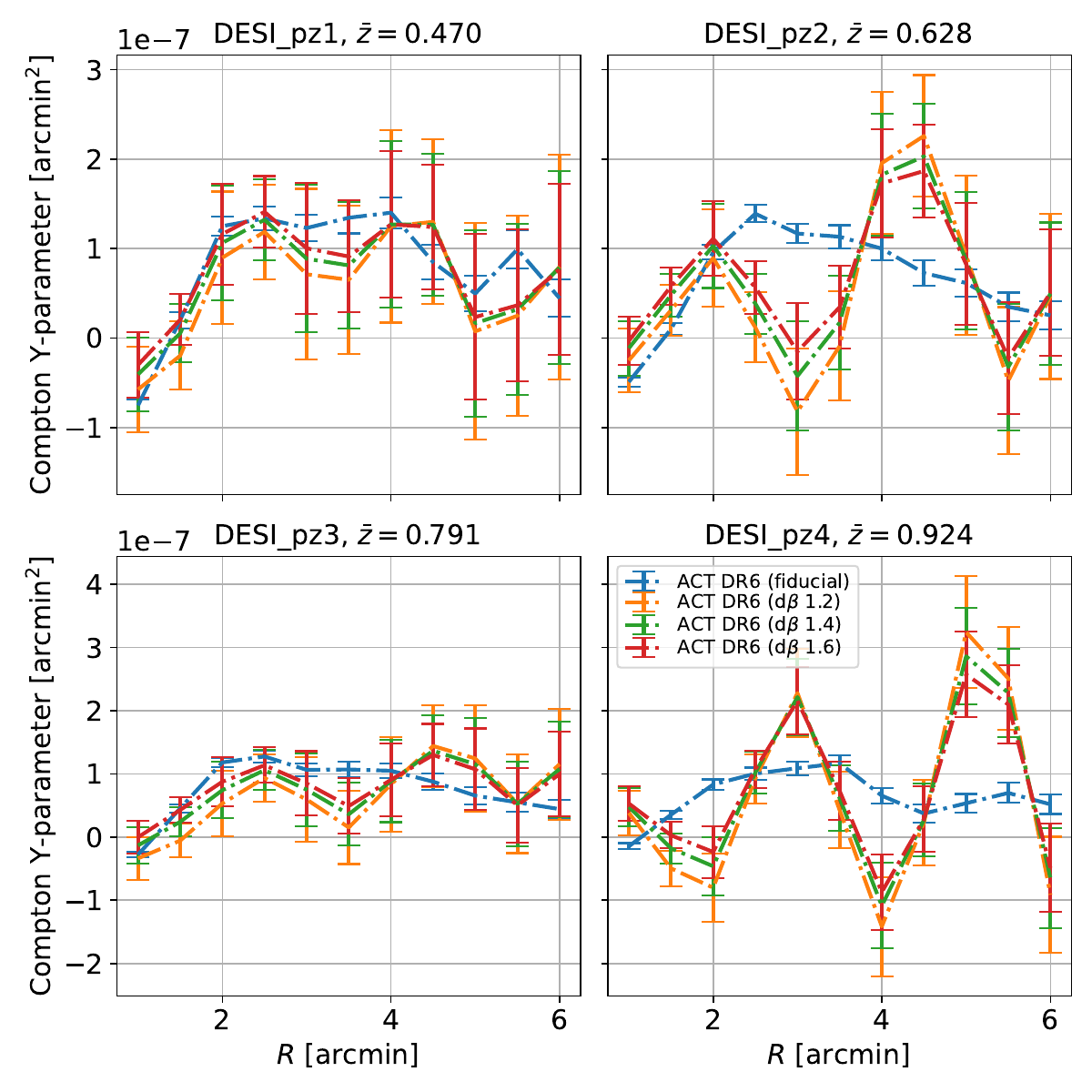}
    \caption{Non-Cumulative variant of Figure \ref{fig:yparam_stackedprofiles_dbeta} showing the stacked profiles for the fiducial and $d\beta$ moment-deprojected CIB-deprojected $y$-parameter maps.}
    \label{fig:stackedProfiles_dBeta_noncum}
\end{figure}

\begin{figure}
    \centering
    \includegraphics[width=\linewidth]{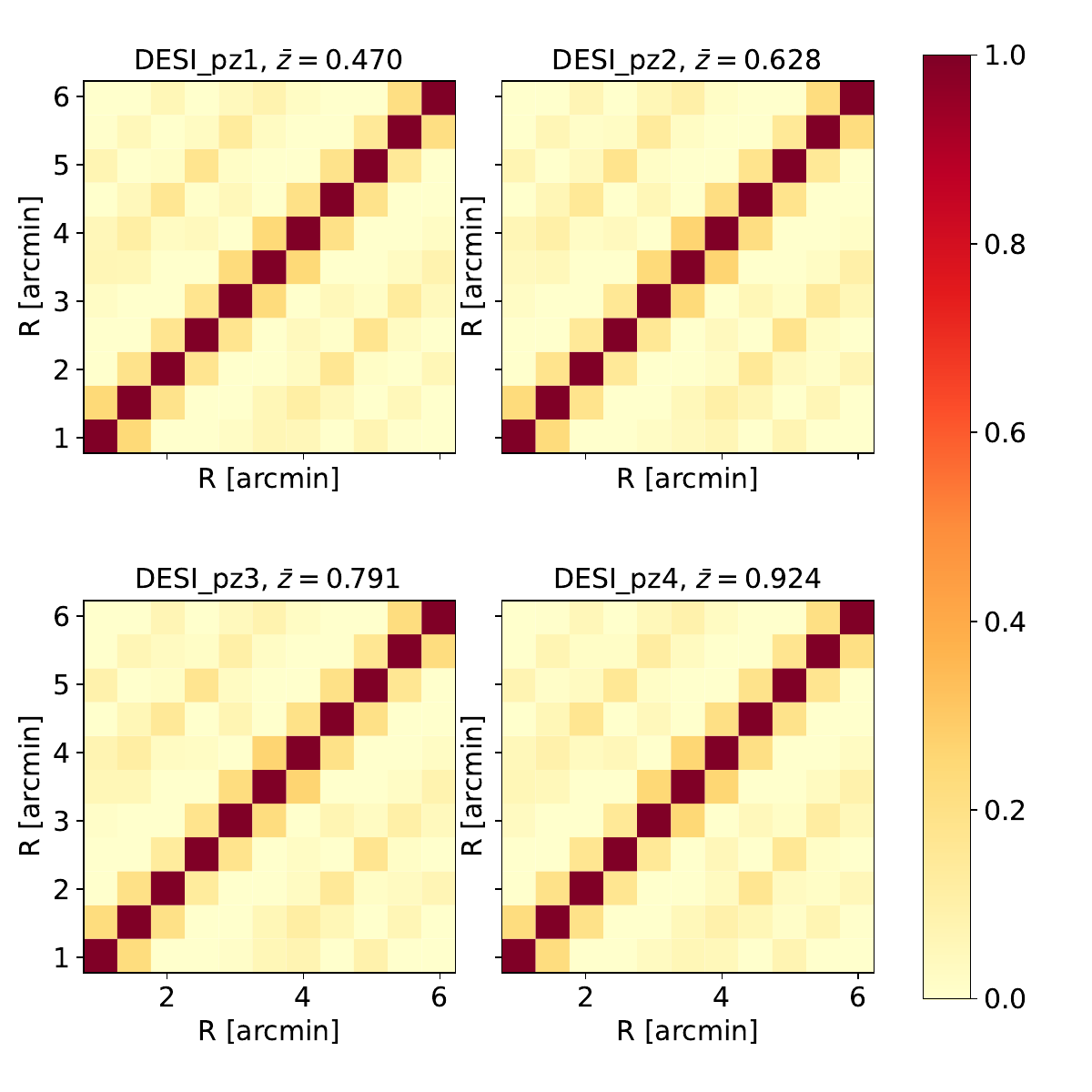}
    \caption{Correlation Matrix for stacking on the fiducial ACT DR6 $y$-parameter map, without CIB deprojection.}
    \label{fig:covariance_matrix_noncum}
\end{figure}

In this appendix, we also present non-cumulative stacked profile measurements, by using a new type of filter we call a ring-ring filter, defined by the following window function:
\begin{equation}
\label{eq:ringring}
    \tilde{W}_{\theta_0, \theta_d}(\theta) = 
    \begin{cases}
     1  & \theta_0 < \theta < \theta_d \\
     -1 & \theta_d < \theta < \sqrt{2\theta_d^2 - \theta_0^2} \\
     0  & \text{otherwise}
    \end{cases}
\end{equation}
for a specified inner radius of $\theta_0$. 
The flexibility of the ring-ring filter is its ability to mask out the central region of a stacked profile, allowing us to measure the outer regions only. In this appendix, the ring-ring filter is used with a varying inner radius to measure the non-cumulative stacked profiles, but as we will see in Appendix \ref{app:Single_Freq}, we can also make use of the ring-ring filter for masking out the central pixels of cluster stacks to study dust effects at large radii.

Using our ring-ring filter as defined above, we can measure the non-cumulative signal at each radius. We do this by fixing the difference between the inner radius $\theta_0$ and $\theta_d$ in Eq. \ref{eq:ringring}  to be a fixed ring width. The outer ring is still maintained at an equal area to the inner ring to cancel out larger fluctuations. For the plots in this section, we pick $R = \theta_d$, and $\theta_0 = \theta_d - 0.5$ arcmin.

The resulting tSZ stacks are shown in Figures \ref{fig:stackedProfiles_noncum}, \ref{fig:stackedProfiles_dBeta_noncum} and \ref{fig:covariance_matrix_noncum}. The non-cumulative nature of this new filter ensures that the covariance matrix is close to diagonal, with the only significant correlation being between neighbouring bins, as shown in Figure \ref{fig:covariance_matrix_noncum}.  

Though we will not be conducting detailed analyses using non-cumulative stacked profiles, we nevertheless present them in this appendix as a possible method for future analysis. 
We note that this non-cumulative stacking is closely related to the slope of the cumulative stacked profiles given previously in this work, and as a result, this is simply a new visualization of the same information. 


\section{Results with $T_{\rm CIB} = 24$K}
\label{App:B}

Alongside the results in the main body, we also present the stacked profile results for a different set of CIB parameters. For example, previous work by \cite{Madhavacheril2020} adopts a different set of CIB parameters with $T_{\rm CIB} = 24\mathrm{K}$ and $\beta=1.2$, which we show below in the following figures for completeness and as a consistency check. 

\begin{figure}[ht]
    \centering
    \includegraphics[width=\columnwidth]{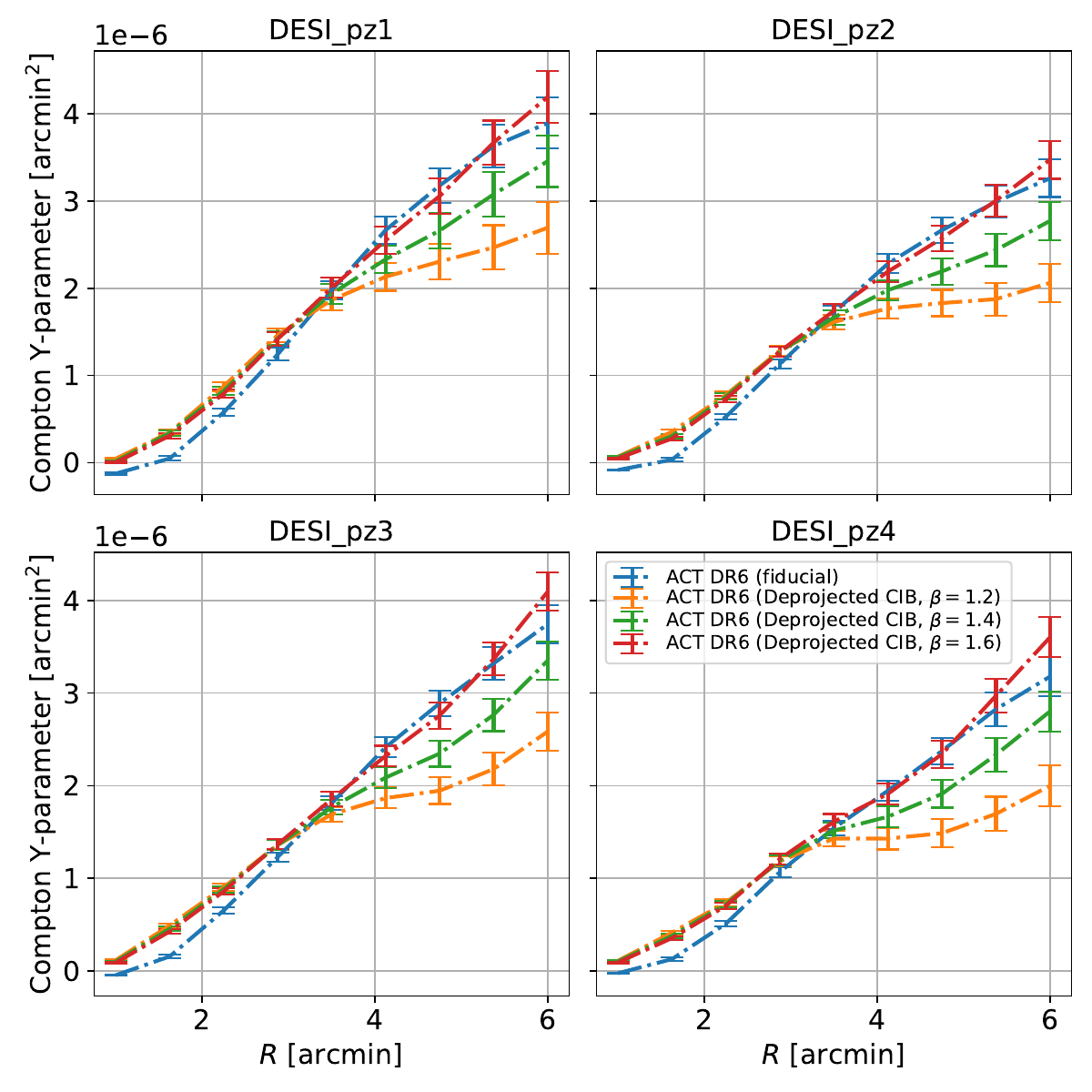}
    \caption{ACT DR6 stacked profiles alongside the deprojected CIB profiles with varying values of $\beta$, for $T_\mathrm{CIB} = 24.0 \mathrm{K}$.}
    \label{fig:yparam_stackedprofiles_cib_24}
\end{figure} 

\begin{figure}[ht]
    \centering
    \includegraphics[width=\columnwidth]{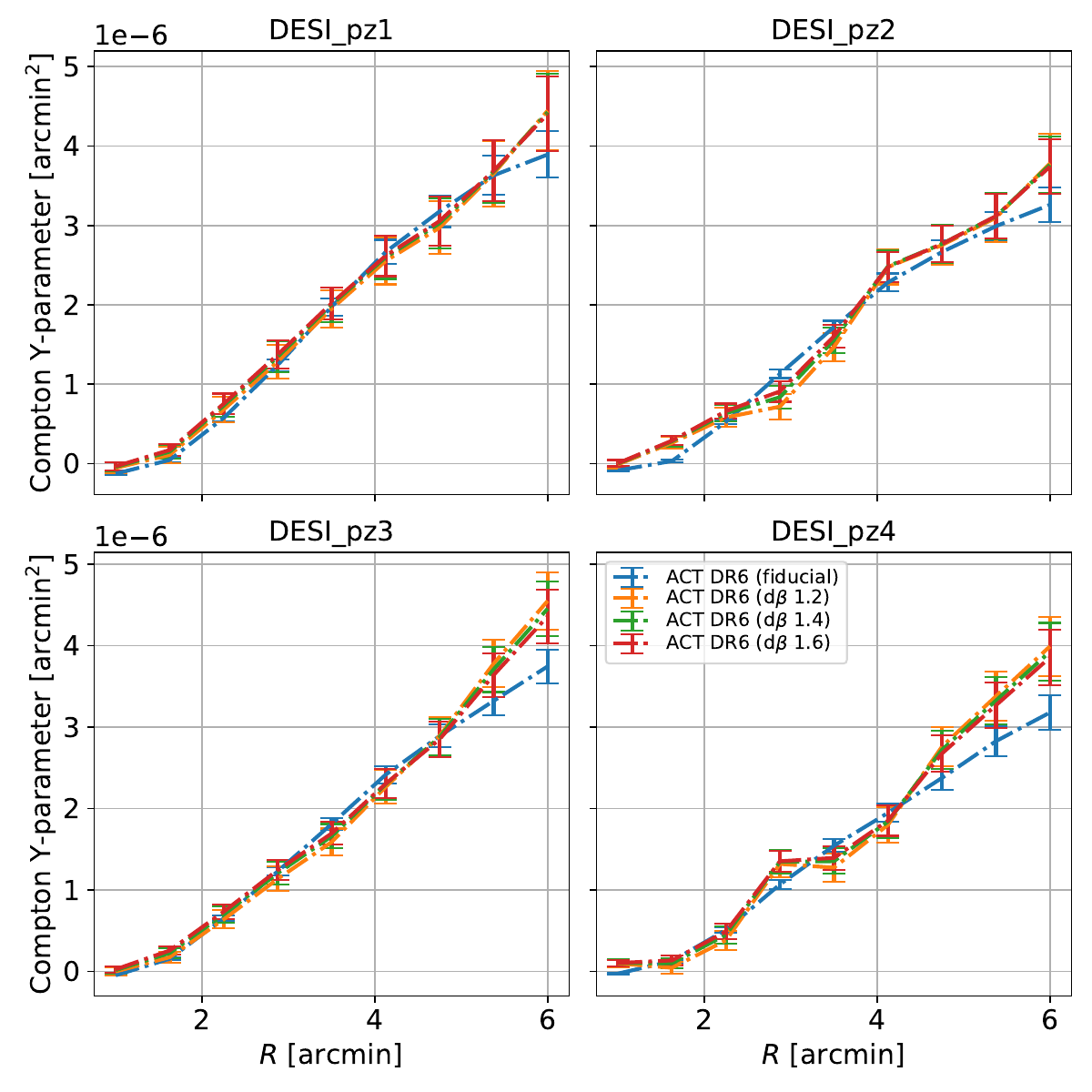}
    \caption{ACT DR6 stacked profiles alongside the deprojected CIB and $d\beta$ profiles, with varying values of $\beta$, for $T_\mathrm{CIB} = 24.0 \mathrm{K}$.}
    \label{fig:yparam_stackedprofiles_dbeta_24}
\end{figure}

In Figure \ref{fig:yparam_stackedprofiles_cib_24}  we can see that for maps with a single $\beta$ deprojection, the results are rather dependent on the fiducial value of $\beta$, a qualitatively similar behaviour as what we have found in Figure \ref{fig:yparam_stackedprofiles_cib} for $T_{\rm CIB} = 10.7$ K.

In Figure \ref{fig:yparam_stackedprofiles_dbeta_24}, we also find that moment deprojection yields results that remain stable despite variations in the fiducial $\beta$ and are consistent across our two choices of $T_{\rm CIB}$. This further reinforces the reliability of our measurements.


\section{Tracking the spatial distribution of dust with single-frequency maps}
\label{app:Single_Freq}


\begin{figure}
    \centering
    \includegraphics[width=\columnwidth]{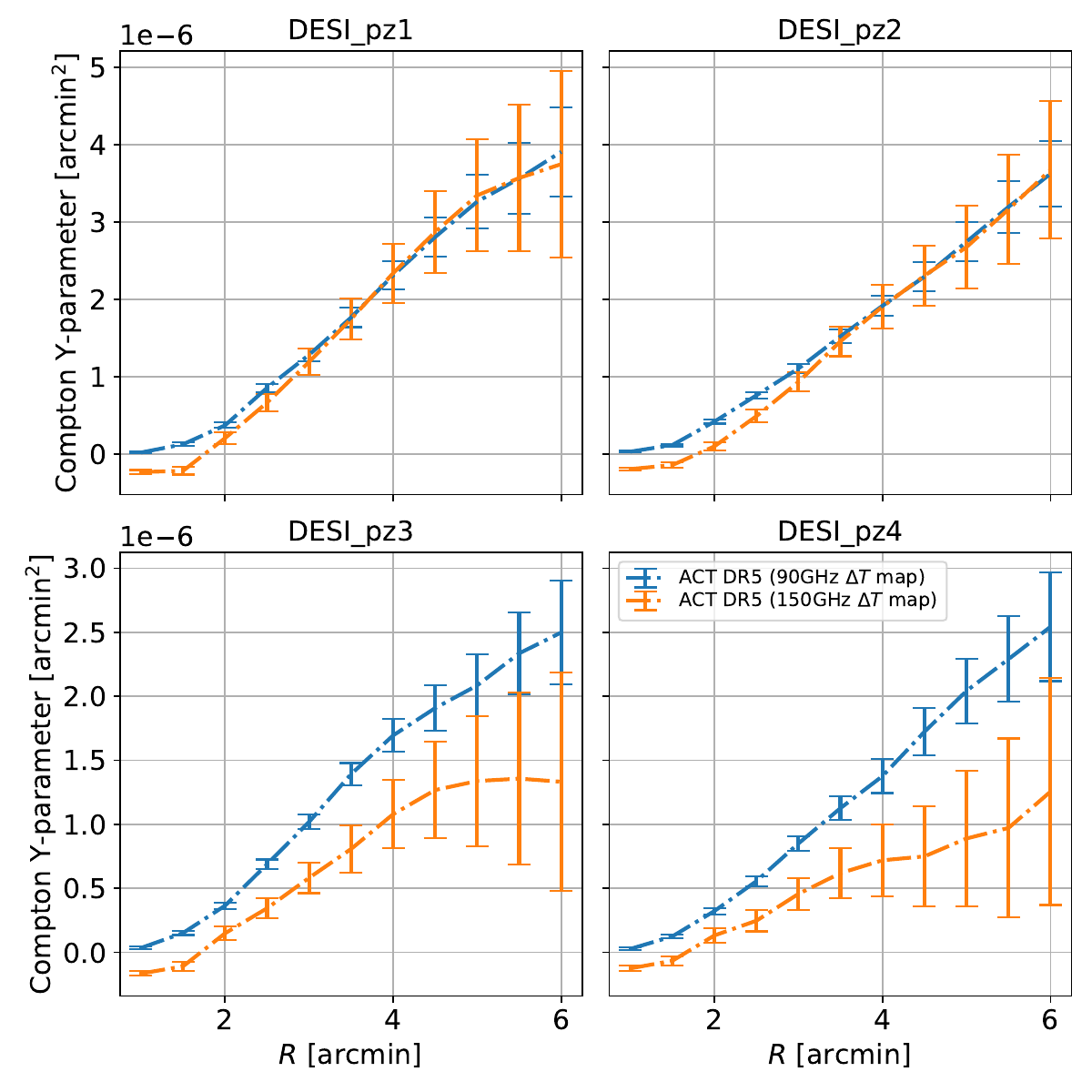}
    \caption{Plot of the ACT DR5 stacked profiles using the standard CAP filter as described by the window function in Eq. \ref{eq:diskring}. Note that the tSZ stacked profiles shown in the main body are all computed with the same filter.}
    \label{fig:singlefreq_diskring}
\end{figure}

\begin{figure}
    \centering
    \includegraphics[width=\columnwidth]{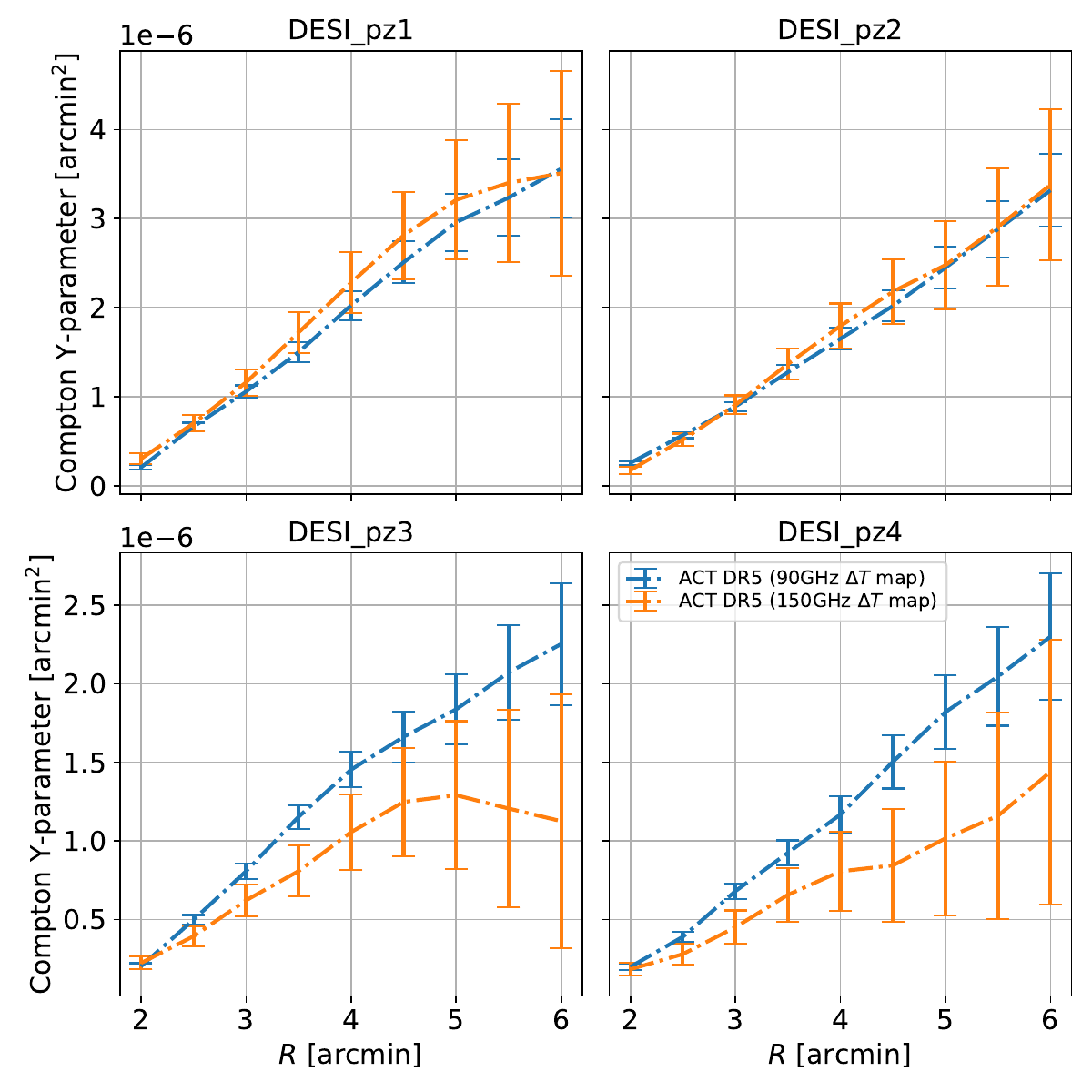}
    \caption{Plot of the ACT DR5 stacked profiles using the ring-ring filter. The filter masks out the central area of the stack following Eq. \ref{eq:ringring}, with $\theta_0=1$ arcmin.}
    \label{fig:singlefreq_ringring}
\end{figure}

To better understand the presence and spatial extent of dust emission in our samples, we also consider stacks on single-frequency maps in this appendix. In particular, we use the ACT DR5 maps \cite{Naess2020_ACTDR5} at the f090  (77–112 GHz) and f150  (124–172 GHz).\footnote{We note that this analysis was conducted predating the creation of ACT DR6 single-frequency maps.}
While they are suboptimal for extracting the tSZ signal and hence not used directly in our fiducial analysis, single-frequency maps maintain the spatial morphology of all components as they are not subject to the same needlet ILC harmonic-space weighting as the processed $y$-parameter maps. Therefore, they allow us to study where the dust emission is localized in space, leading to a better understanding of the CIB contamination. 

In particular, we expect that a large fraction of the dust emission is associated with the very centre of these galaxies, where the stars reside. This is because it is the stars that both create the dust grains and heat them up, leading to infrared emission. To test this hypothesis, we compare the temperature decrement in the f090 and f150 maps converted to units of Compton-$y$, using Eq. \ref{eq:tSZ}. We note that the tSZ signal should be the same, but the dust signal should be a factor of several larger in f150 compared to f090. Therefore, we will attribute any statistical difference between the f150 and f090 stacks to dust.

There is one further complication: since our CAP filters are cumulative, a difference in the central pixel will reflect as a constant offset between the two curves. 
However, we can solve this problem using the ring-ring filter we defined in Appendix \ref{app:noncumulative}, which we can use to explicitly remove the central region where we expect most of the dust emissions to originate.


Following the ring-ring filter given in Eq. \ref{eq:ringring}, we choose the inner radius $\theta_0 = 1$ arcmin to mask the thermal dust emissions otherwise contaminating the tSZ signal at the centre of the stacked profile. 
As in the case of the standard CAP filter, the ring-ring filter's outer radius is chosen to maintain an equal area as the inner ring so that the same filtering out of large-scale fluctuations of the CMB also applies here. 


Comparing the standard CAP filter stacks from Figure \ref{fig:singlefreq_diskring} and the ring-ring filtered stacks from Figure \ref{fig:singlefreq_ringring}, we can easily observe the effects of dust emissions at the centre of the f150 band in the CAP measurements: this shows up as a rather significant detection of ``negative'' $y$ parameter in f150, likely indicating a positive emission in the f150 map at the centre of the stack.

Interestingly, for the ring-ring filter the lowest two redshift bins pz1 and pz2, which are also the most massive, we find excellent consistency at all apertures, strongly suggesting that the only significant dust contamination is associated with the stellar disk \footnote{Which at our resolution is essentially a ``delta function'' at the center of the galaxy, but convolved with the 1.4 arcmin beam for f150 and 2.1 arcmin for f090.} in the innermost apertures. This should be compared to the behaviour of the CIB deprojected stacks shown in Figure \ref{fig:yparam_stackedprofiles_cib}, which show large differences between the fiducial curve and the CIB-deprojected ones at large apertures. We speculate that the harmonic filtering applied to the maps through the NILC algorithm does not preserve the spatial morphology of the dust emission and ``spreads'' it to all apertures. 

On the contrary, we find some mild hints of dust contamination at larger apertures for pz3 and pz4, due to the difference of the stacked profiles at the two frequencies. Similar hints have also been noted in  \cite{Menard2010}. These galaxies are at higher redshift, closer to the peak in star formation, and have a lower mass, meaning that the two-halo term of the dust (i.e. dust in other galaxies correlated to the ones of interest) is expected to be larger. Alternatively, it could indicate that dust has been pushed to the outskirts of the halos by feedback mechanisms that lead to an extended dust profile. While not the main focus of this work, this interesting hint warrants further study.

\begin{figure}
    \centering
    \includegraphics[width=\columnwidth]{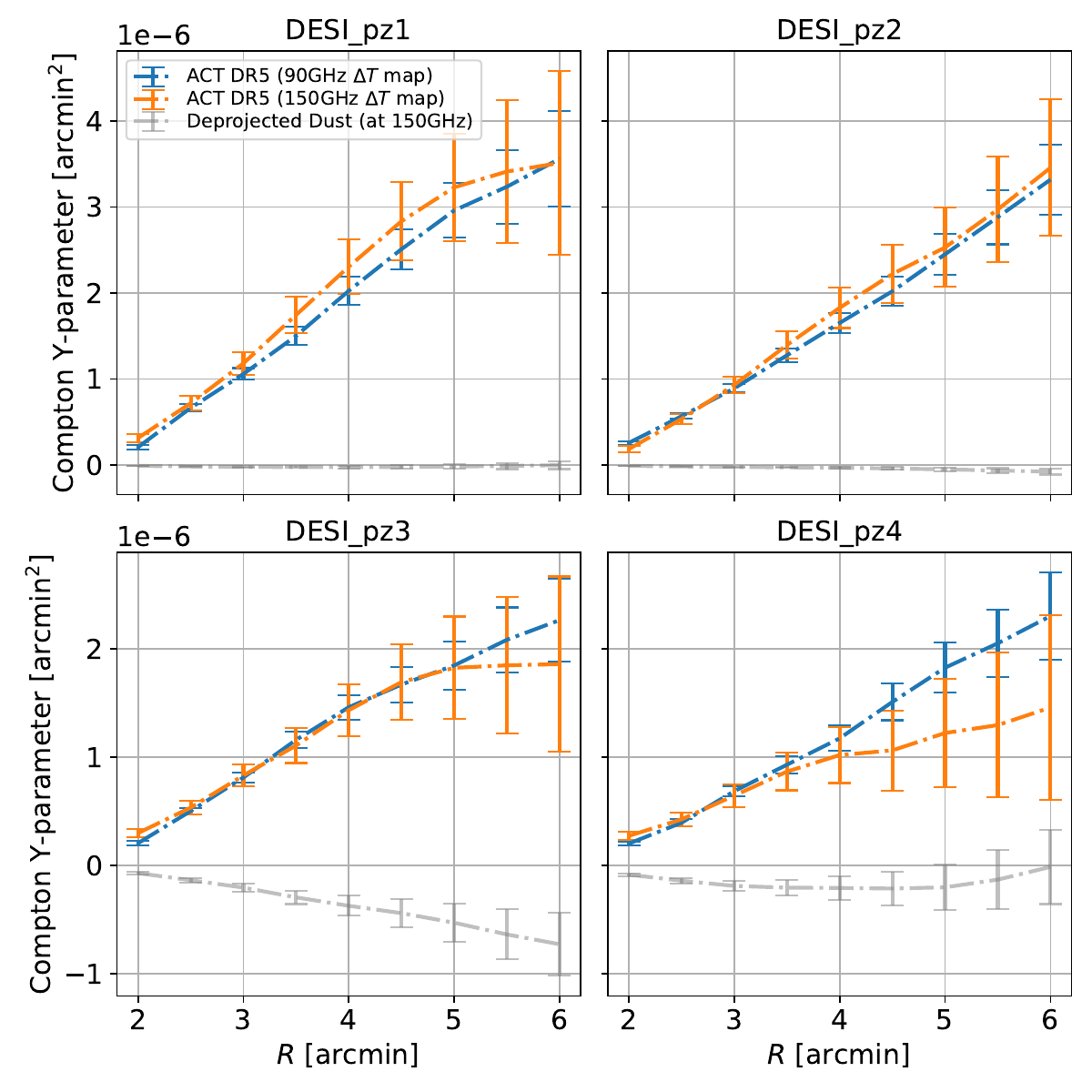}
    \caption{Plot of the ACT DR5 stacked profiles with deprojected dust, using the ring-ring filter.}
    \label{fig:singlefreq_deprojected}
\end{figure}

We can further study the potential presence of dust in the outskirts of the higher redshift bins by the f220 CMB Map from ACT. As the tSZ effect is null at 218 GHz, stacking on the f220 map isolates the dust signal with minimal to no contributions from the tSZ. We can then rescale the f220 dust signal to other map frequencies using the modified CIB Blackbody frequency dependence, as defined in Eq. \ref{eq:f_CIB}. By treating the spectral index $\beta$ in Eq. \ref{eq:f_CIB} as a free parameter for each sample (since we have argued that we have a large uncertainty on $\beta$), we can then minimize\footnote{This is done by minimizing the $\chi^2(\beta)$ between the two curves, accounting for the proper covariance matrices. This is equivalent to jointly fitting the tSZ and dust profiles, assuming that there is no tSZ signal in the f220 map.} the difference between the 90 and 150 GHz stacks when correcting for the dust profile rescaled to the respective frequencies. The best-fit dust signal, extrapolated to 150 GHz, is shown in light grey in Figure \ref{fig:singlefreq_deprojected}. We choose to apply this fit to the ring-ring filtered stacked profiles, rather than the disk-ring filtered versions, as our goal is to model the CIB contamination at large angular scales. 


As we can see, the best-fit dust profiles (extrapolated to 150 GHz) in the first two photo-z bins are consistent with null, in agreement with our earlier conclusions. However, we do see moderate evidence of dust in the outskirts of pz3 and pz4, as shown by the grey curves in the bottom panels of Figure \ref{fig:singlefreq_deprojected}. This again is in line with earlier conclusions. The origin of this extended dust emission, and whether it is associated with the galaxies themselves or originates from other correlated galaxies (2-halo term), is an interesting question and worthy of further study. 




\end{document}